\documentclass{elsart}
\usepackage{epsfig}
\usepackage{amstex}
\usepackage{supertabular}

\newlength{\la}
\newlength{\lb}
\newlength{\lc}
\newlength{\pa}
\newlength{\pbb}
\newlength{\pcc}
\newlength{\pdd}

\newcommand{\file}{\sf}
\newcommand{\cpindent}[1]{\parbox[t]{\lc}{\raggedright \begin{tabular}{lp{\pa}} #1 \end{tabular}}}

\newcommand{\lindent}[1]{\parbox[t]{\lc}{\hspace*{-0.25cm}\begin{tabular}{p{1.5cm}p{7.1cm}} #1 \end{tabular}}}
\newcommand{\lindentt}[1]{\parbox[t]{\pbb}{\hspace*{-0.25cm}\begin{tabular}{p{0.5cm}p{4.0cm}} #1 \end{tabular}}}
\newcommand{\tspc}{\hspace*{0.3cm}}

\newcommand{\immath}{\mathrm{i}}

\begin{document}
\psfigdriver{dvips}
\begin{frontmatter}
\title{Density-functional theory calculations for poly-atomic systems:
Electronic structure, static and elastic properties and {\em ab initio}
molecular dynamics}
\author{Michel Bockstedte\thanksref{MBockstedte},}\author{Alexander Kley, J\"org Neugebauer and Matthias Scheffler}
\address{ Fritz-Haber-Institut der Max-Planck-Gesellschaft, Faradayweg 4-6,\\
 D-14195 Berlin-Dahlem, Germany} 
\thanks[MBockstedte]{Present address: Lehrstuhl
f. theor. Fest\"orperphysik, Universit\"at Erlangen-N\"urnberg,
Staudtstr. 7/B2, D-91058 Erlangen}
\begin{abstract}
The package {\sf fhi96md} is an efficient code to perform density-functional
theory total-energy calculations for materials ranging from insulators to
transition metals. The package employs 
first-principles pseudopotentials, and a plane-wave basis-set.
For exchange and correlation both
the local density and generalized gradient approximations are implemented.  The
code has a low storage demand and performs efficiently on low budget
personal computers as well as high performance computers.
\end{abstract}

\end{frontmatter}
{
\twocolumn
\section*{PROGRAM SUMMARY}
{\em Title of the Program:\quad} {\sf fhi96md}\\[1ex]
{\em catalogue number:\quad} \\[1ex]
{\em Program obtainable from:\quad} CPC Program Library, Queen's University of
Belfast, N. Ireland\\[1ex]
{\em Licensing provisions:\quad} Persons requesting the program must sign the
standard CPC non-profit use license (see license agreement printed in every
issue).\\[1ex] 
{\em Computer for which the program is designed and others on which it has
  been tested:\quad} IBM RS/6000, Pentium-PC\\[1ex]
{\em Installation:\quad} Fritz-Haber-Institut der Max-Planck-Gesellschaft,
Berlin-Dahlem\\[1ex]
{\em Operating system:\quad} UNIX\\[1ex]
{\em Programming language:\quad} FORTRAN 77\\[1ex]
{\em Memory required to execute with typical data: \quad} 1-32 Mwords\\[1ex]
{\em No.~of bits in a word:\quad} 32\\[1ex]
{\em Memory required for test run:\quad} 0.5 MB\\[1ex]
{\em Time for test run\/}:\quad 6\,min\\[1ex]
{\em No. of lines in distributed program, including test data, etc.:\quad}
14\,000\\[1ex]

{\em Keywords:\quad} 
density-functional, local-density, generalized gradient,
pseudopotentials, plane-wave basis, super cell, molecular dynamics,  structure 
optimization, total-energy, potential-energy surfaces, chemical binding, diffusion,
surface reactions, crystals, defects in crystals, surfaces, molecules\\[1ex]

{\em Nature of the physical problem\quad} \\[1ex] In poly-atomic systems as
for example mole\-cu\-les~\cite{andreoni:92,troullier:92},
crystals~\cite{ortiz:92,gracia:92}, defects in
crystals~\cite{dabrowski:89,zhang:91},
surfaces~\cite{alves:91,neugebauer:92,stumpf:96}, it is highly desirable to
perform accurate electronic structure calculations, without introducing
uncontrollable approximations. This enables the predictive description of
equilibrium properties as well as of non-equilibrium phenomena for a wide
class of materials. Examples studied with the present code or its predecessor
include meta-stabilities of defects \cite{dabrowski:89,caldas:90}, surface
reconstructions~\cite{pehlke:93,pankratov:95},
diffusion \cite{stumpf:96,bockstedte:96}, surface
reactions~\cite{gross:94,pehlke:95,stampfl:96}, and phase
transitions~\cite{moll:95}.  Molecular dynamics simulations combined with
first-principles forces are a powerful tool to analyze the motion of the
nuclei~\cite{bockstedte:96,gross:96} and to accurately calculate thermodynamic
properties such as diffusion constants and free energies~\cite{bockstedte:96}.

The computer code described below enables this variety of investigations. It
employs density-functional theory~\cite{hohenberg:64} together with the
local-density approximation~\cite{ceperley:80,perdew:81} or generalized
gradient approximations~\cite{becke:88,perdew:86,perdew:92} for the
exchange-correlation functional.

{\em Method of solution}\\[1ex] {\em Ab initio} molecular dynamics on the
Born-Oppen\-heimer surface is implemented by a two step method.  In the first
step the Kohn-Sham equation~\cite{kohn:65} is solved self-consistently to
obtain the electron ground state and the forces on the
nuclei. In a second step these forces are used to integrate the equations of
motion for the next time step.  The calculation of the total-energy and the
Kohn-Sham operator in a plane-wave basis-set is done by the momentum-space
method~\cite{ihm:79}. To solve the Kohn-Sham equation the package {\sf
  fhi96md} employs the iterative schemes of Williams and Soler~\cite{williams:87}
and Payne {\em et~al.}~\cite{payne:86}. We use the frozen-core approximation
and replace the effect of the core electrons by norm-conserving
pseudopotentials~\cite{bachelet:82,hamann:89,troullier:91,gonze:91} in the
fully separable form~\cite{kleinman:82}. The equations of motion of the nuclei
are integrated using standard schemes in molecular dynamics such as the Verlet
algorithm. Optionally an efficient structure optimization can be performed by
a second order algorithm with a damping term.

{\em Restrictions on the complexity of the problem\quad}\\[1ex] Only
pseudopotentials with $s$-, $p$-, and $d$-compo\-nents are implemented.  For
highly correlated systems (e.g. $f$-electrons) the treatment of the
exchange-correlation interaction is not appropriate.  Relativistic effects are
included via scalar relativistic pseudopotentials.  The system is assumed
to be non-magnetic, but a generalization of the program to magnetic states is
straight forward.\\[1ex] }

\onecolumn
{\bf \indent LONG WRITE-UP}
\section{Introduction} Total-energy calculations and molecular dynamics
simulations employing density-functional theory~\cite{hohenberg:64} represent a
reliable tool in condensed matter physics, material science, chemical physics
and physical chemistry. A large variety of applications in systems
as different as molecules~\cite{andreoni:92,troullier:92}, bulk
materials~\cite{ortiz:92,gracia:92,dabrowski:89,zhang:91} and
surfaces~\cite{alves:91,neugebauer:92,stumpf:96} have proven the power of
these methods in analyzing as well as in predicting equilibrium and
non-equilibrium properties. {\em Ab initio} molecular dynamics
simulations enable the analysis of the atomic motion and allow the accurate
calculation of thermodynamic properties such as the free energy, diffusion
constants and melting temperatures of materials.

The package {\file fhi96md} described in this paper is especially designed to
investigate the material properties of large systems. It is based
on an iterative approach to obtain the electron ground state. Norm-conserving
pseudopotentials~\cite{bachelet:82,hamann:89,troullier:91,gonze:91} in the
fully separable form of Kleinman and Bylander~\cite{kleinman:82} are used to
describe the potential of the nuclei and core electrons  acting on the valence electrons. Exchange and
correlation are described by either the local-density
approximation~\cite{ceperley:80,perdew:81} or generalized gradient
approximation~\cite{becke:88,perdew:86,perdew:92}. The equations of motion of
the nuclei are integrated using standard schemes in molecular dynamics.
Optionally an efficient structure optimization can be performed by a damped
dynamics scheme.

The package {\sf fhi96md} is based on a previous version {\sf
  fhi93cp}~\cite{stumpf:94}. Advances of the new version are an
improved iteration scheme to solve the Kohn-Sham equations, the generalized
gradient approximation for exchange and correlation, a mixed basis-set
initialization and molecular dynamics. The package consists of the
program {\sf fhi96md} and a start utility {\sf start}.  The program {\sf
  fhi96md} can be used to perform static total energy calculation or {\em ab initio}
molecular dynamics simulation.  The utility {\sf start} assists in generating
the parameter file and the input file required to compile and run {\sf
  fhi96md}, thereby utilizing a low memory demand for each individual run.

The paper is organized as follows. In Section~\ref{sec:moldyn} we briefly
outline the method as implemented in the package. Section~\ref{sec:impl}
describes the program structure and the input and output files of the program.
The last two Sections concern the installation of the package and the test
run. Input and output of the test run are found at the end of the paper.
Appendix~\ref{sec:expr} lists the formulas and expressions
implemented in the package and Appendix~\ref{sec:app_tables} describes the
parameter and input files as generated by the start utility.

\section{{\em Ab initio} molecular dynamics}
\label{sec:moldyn}
The key approach to molecular dynamics (MD) is $(i)$ the separation of the slow
motion of the nuclei from the fast motion of the electrons within the
Born-Oppenheimer approximation and $(ii)$ that the motion of the nuclei is
governed by Newton's equation of motion:
\begin{equation}
\label{eq:md_bo}
M_J\,\frac{d^2}{{d\/t}^2}{\bf R}_J=-\frac{\partial}{\partial\/{\bf R}_J}
E_0\left(\{{\bf R}_J\}\right)
\end{equation} where $M_J$ and ${\bf R}_J$ are the masses and coordinates of
the $N_{\rm at}$ atoms rsp. and $E_0\left(\{{\bf R}_J\}\right)$ is the many-electron
ground state energy.  We employ density-functional theory (DFT) together with
common approximations to the exchange-correlation energy functional to
obtain the ground state electron density and the forces acting on the nuclei.
Computationally the ground state electron density is obtained by
self-consistently solving the Kohn-Sham equations in a pseudopotential
plane-wave approach. Once the forces acting on the nuclei are calculated the
equations of motion are integrated numerically. To perform
simulations in the canonical ensemble a Nos\'{e}-Hoover
thermostat~\cite{nose:84,hoover:85} is employed to control the temperature
during the simulation.  The methods involved in finding the electronic ground
state and integrating the equation of motion of the nuclei are outlined in the
first two Subsections.

Frequently one is only interested in the geometry of stable and
metastable structures, e.g. of surfaces, defects, or complicated crystals. The
equilibrium geometry and the corresponding total-energy can be obtained by
relaxing the coordinates of the nuclei starting from a guessed geometry.  An
example for this class of application is the calculation of
adiabatic potential energy surfaces, which are widely used to study
surface reactions or defect migration on the microscopic level. An outline of
the techniques for structure optimization is given at the end of this
Section.

\subsection{Solving the Kohn-Sham equations}
\label{sec:elproblem}
\paragraph{The energy functional} The key variable in DFT is the electron 
density $n({\bf r})$. As stated by the fundamental theorem of Hohenberg and
Kohn~\cite{hohenberg:64} the ground state energy $E_0(\{{\bf R}_J\})$ of the
system for given positions of the nuclei $\{{\bf R}_J\}$ is the minimum of the
Kohn-Sham total energy functional~\cite{kohn:65} with respect to the electron
density $n$. The total energy functional $E\left[n\right]$ is:
\begin{equation}
\label{eq:Etot}
E\left[n\right]=T^{\rm s}\left[n\right]+E^{\rm H}\left[n\right]
+E^{\rm e-nuc}\left[n\right]+E^{\rm XC}\left[n\right]+E^{\rm nuc-nuc}\quad,
\end{equation} where $T^{\rm s}$ is the kinetic energy of non-interacting
electrons, $E^{\rm H}$ is the Hartree energy, and $E^{\rm XC}$ is the
exchange-correlation energy. The energy expressions are briefly discussed in
the following. The explicit expressions for each of the contributions to the
total-energy, potentials and forces as implemented in the program can be found
in Appendix~\ref{sec:expr}.

The energy of the electron-nuclei and nuclei-nuclei interaction $E^{\rm
e-nuc}$ and $E^{\rm nuc-nuc}$ are
\begin{equation}
\label{eq:Eion}
E^{\rm e-nuc}\left[n\right]=\int d^3\/r\,V^{\rm e-nuc}({\bf r})\/n({\bf r})\quad\quad
{\rm and}\quad\quad
E^{\rm nuc-nuc}=\frac{1}{2}\sum_{I\/J,\,I\neq J} \frac{Z_I\,Z_J}{|{\bf R}_I-{\bf R}_J|}\quad,
\end{equation}
where $Z_I$ and $Z_J$ are the charges of the corresponding
nuclei. Throughout the paper we employ atomic units (energy in hartree) unless
otherwise noted.  As approximations to the exchange-correlation energy
functional $E^{\rm XC}\left[n\right]$ we employ the local-density
approximation (LDA) -- as obtained from the homogeneous electron gas by
Ceplerley and Alder~\cite{ceperley:80} in the parameterization of Perdew and
Zunger~\cite{perdew:81} -- and the generalized gradient approximations of
Becke and Perdew~\cite{becke:88,perdew:86} (BP), and of Perdew {\em et~al.}~\cite{perdew:92} (PW91).

The system is represented by the super cell approach.
The geometry of the nuclei is contained in a super cell, which is periodically
repeated on a lattice. The coordinates ${\bf R}_J$ of a nucleus or its
periodic image hence are 
\[
{\bf R}_J=\boldsymbol{\tau}_{I_{\rm s},I_{\rm a}}+{\bf R}\quad\quad{\rm
with}\quad J=\{I_{\rm
s},I_{\rm a},{\bf R}\}\quad,
\] where the index
$I_{\rm s}$ for the species of a nucleus, the index $I_{\rm a}$ for the
nucleus itself and the lattice vector ${\bf R}$ pointing to the
origin of the cell or of its image.  

The effect of the core electrons and
the coulomb potentials of the nuclei is replaced by soft pseudopotentials
which enables the efficient use of a  plane-wave basis:
\begin{equation}
\hat{V}^{e-nuc}=\sum_{\bf R}\sum_{I_{\rm s},I_{\rm a}}\hat{V}^{\rm
e-nuc}_{I_{\rm s}}({\bf r}-{\bf \tau}_{I_{\rm s},I_{\rm a}}-{\bf R})
\end{equation} We employ norm conserving pseudopotentials constructed e.g.
following the schemes of Hamann~\cite{hamann:89} or Troullier and
Martins~\cite{troullier:91}. These have proven to yield 
transferable potentials for a broad class of nuclei ranging from first row
elements to transition metals.
The pseudopotentials are represented in the fully separable form as proposed by
Kleinman and Bylander~\cite{kleinman:82}
\begin{equation}
\label{eq:ps}
\hat{V}^{\rm e-nuc}_{I_{\rm s}}=\hat{V}_{I_{\rm s},l_{\rm loc}}+
\sum_{\/^{l=0,}_{l\neq l_{\rm loc}}}^{l_{\rm max}}\sum_{m=-l}^{l}
\frac{|\Delta V_{I_{\rm s},l}^{\rm nl}|\psi^{\rm ps}_{I_{\rm s},l,m}\rangle
\langle\psi^{\rm ps}_{I_{\rm s},l,m}|\Delta V_{I_{\rm s},l}^{\rm nl}|}
{\langle \psi^{\rm ps}_{I_{\rm s},l,m}|\Delta V_{I_{\rm s},l}^{\rm nl}|\psi^{\rm ps}_{I_{\rm s},l,m}\rangle}\quad,
\end{equation} where $\Delta V_{I_{\rm s},l}^{\rm nl}(r)=V_{I_{\rm
    s},l}(r)-V_{I_{\rm s},l_{\rm loc}}(r)$, and $V_{I_{\rm s},l}(r)$ are the
radial components of the semi-local pseudopotential and $\psi^{\rm ps}_{I_{\rm
    s},l,m}({\bf r})=R^{I_{\rm s}}_l(r)\/Y_{l}^{m}(\theta,\phi)$ are the
node-free atomic pseudo wave functions. In this form the pseudopotential is
splitted into a local part $\hat{V}^{\rm ps,local}_{I_{\rm s}}$ and a non-local but
separable part $\hat{V}^{\rm ps,nl}_{I_{\rm s}}$.  Correspondingly the potential $\hat{V}^{\rm
  e-nuc}$ and the energy $E^{\rm e-nuc}$ are expressed as
\begin{equation}
\hat{V}^{\rm e-nuc}=\hat{V}^{\rm ps,local}+\hat{V}^{\rm ps,nl}
\quad\quad{\rm and}\quad\quad E^{\rm e-nuc}=E^{\rm ps,local}+
E^{\rm  ps,nl}\quad.
\end{equation} 

For some atomic species such as sodium, zinc and copper, it is essential to
include the non-linear core valence exchange-correlation description
(NLCV-XC)~\cite{hebenstreit:91,louie:82}. In this case the effect of the core
electron density on the exchange-correlation energy is described by a pseudo
core density $\tilde{n}^{\rm core}({\bf r})$, which is the superposition of
the smoothed core densities $\tilde{n}^{\rm core}_{I_{\rm s}}( |{\bf
r}-\boldsymbol{\tau}_{I_{\rm s},I_{\rm a}}-{\bf R}|)$ constructed together
with the pseudopotentials~\cite{fuchs:96}.  The exchange-correlation
functionals $E^{\rm XC}\left[n\right]$ and $V^{\rm XC}\left[n\right]$ are then
replaced by $E^{\rm XC}\left[n+\tilde{n}^{\rm core}\right]$ and $V^{\rm
XC}\left[n+\tilde{n}^{\rm core}\right]$.

\paragraph{Kohn-Sham equations} In the Kohn-Sham scheme~\cite{kohn:65} the
electron density is expressed by a set of orthogonal, normalized Kohn-Sham orbitals
$\phi_\alpha({\bf r})$ 
\begin{equation}
\label{eq:eldens}
n({\bf r})=\sum_{\alpha}\/f_{\alpha}\,|\Phi_{\alpha}({\bf r})|^2.
\end{equation} The occupation numbers $f_{\alpha}$ vary between $0$ and $2$ as
the electron spin is not included explicitly and the sum over all occupation
numbers is the total number of electrons $N_{\rm el}$ per super cell. The
ground state electron density is calculated by solving the Kohn-Sham equations
self-consistently for these orbitals
\begin{equation} 
  \underbrace{ \left(-\frac{1}{2}\nabla^2+{V}^{\rm e-nuc}+{V}^{\rm
      H}\left[n\right]+{V}^{\rm XC}\left[n\right] \right)}_{\hat{H}^{\rm
      KS}}\,\phi_{\alpha}({\bf r})=\epsilon_{\alpha}\,\phi_{\alpha}({\bf r})\quad.
\end{equation} This is equivalent to a constrained minimization of the total energy functional
$E[n]$ with respect to the Kohn-Sham orbitals 
\begin{equation}
E_{0}= \min_{\{|\phi_{\alpha}\rangle\}} E[n]\quad\quad{\rm with}\quad\quad
\langle \phi_{\alpha}|\phi_{\beta}\rangle=\delta_{\alpha\/\beta}\quad\quad{\rm
and}\quad\quad\int n({\bf r})\,d^3r=N_{\rm el}.
\end{equation} 

\paragraph{Occupation numbers} The occupation numbers $f_{\alpha}$ are 
determined by the Fermi distribution
\begin{equation} f_{\alpha}=2\,\frac{1}{e^{\frac{\epsilon_{\alpha}-\mu_f}{k_B\/T_{\rm
        el}}}+1},
\end{equation} where the Fermi energy $\mu_f$ is defined by the number of
electrons $N_{\rm el}=\sum_{\alpha}f_{\alpha}$ and $T_{\rm el}$ is the
temperature.  It is convenient to use an artificially high temperature (e.g.
$k_{\rm B}\/T_{\rm el}=0.1\,{\rm eV}$), since a broadening of the occupation
improves the stability and the speed of convergence of the calculation. However, as pointed out
in Ref.~\cite{weinert:92} this corresponds to minimizing the free energy
functional \mbox{$F[n]=E[n]-T_{\rm el}\,S_{\rm el}$} with the entropy \mbox{$S_{\rm
el}=\sum_\alpha (f_\alpha {\rm ln}\/f_\alpha+(1-f_\alpha){\rm
ln}(1-f_\alpha))$} rather than the total-energy $E[n]$.  In order to obtain the
total-energy at $T_{\rm el}=0\,{\rm K}$, we employ that
\mbox{$E_0=E[n]-\frac{1}{2}\/T_{\rm el}\,S_{\rm el}$} yields the total-energy at
$T_{\rm el}=0\,{\rm K}$ up to $O(T_{\rm
el}^3)$~\cite{neugebauer:92,gillan:89}. However, for typical values of $k_{\rm
B}\/T_{\rm el}$ the deviations are small and the forces on the nuclei remain
accurate~\cite{wagner:96}.

\paragraph{Plane-wave basis-set} The Kohn-Sham orbitals are represented by a plane-wave basis-set
\begin{equation}
\phi_{i,{\bf k}}({\bf r})=\sum_{{\bf G},\frac{1}{2}|{\bf G}+{\bf k}|^2\leq E_{\rm cut}} 
c_{i,{\bf G}+{\bf k}}\,{\rm e}^{\immath\/({\bf G}+{\bf k})\cdot{\bf r}}
\end{equation} truncated at a energy cut-off $E_{\rm cut}$. The Brillouin zone
integral over the k-points is replaced by a sum over a set of special k-points
with the corresponding weights $w_{\bf k}$~\cite{monkhorst:76}. In
the start utility the Monkhorst-Pack scheme~\cite{monkhorst:76} has been
implemented to generate special k-points and for each k-point set the quality
check proposed by Chadi and Cohen~\cite{chadi:73} is automatically performed.

\paragraph{Solving the Kohn-Sham equation}
In the past few years iterative techniques have become the method of choice to solve the
Kohn-Sham equations and enabled first-principles studies even for large
systems. The key idea is to minimize the energy functional with respect to the
wave function $|\phi_{i,{\bf k}}\rangle$ starting with a trial wave function
$|\phi_{i,{\bf k}}^{(0)}\rangle$. The energy minimization scheme is formulated
in terms of an equation of motion (EOM) for the wave function $|\phi_{i,{\bf
    k}}^{(\tau)}\rangle$. The simplest scheme is the steepest descent approach
\begin{equation}
\label{eq:steepest_descent}
\frac{d}{d\/\tau}| \phi_{i,{\bf
k}}^{(\tau)}\rangle=\left(\tilde{\epsilon}_{i,{\bf k}}-\hat{H}_{\rm KS}\right)
|\phi_{i,{\bf k}}^{(\tau)}\rangle\quad,
\end{equation} under the ortho-normality constraint $\langle\phi_{i,{\bf
    k}}^{({\tau})}|\phi_{j,{\bf
    k}}^{({\tau})}\rangle=\delta_{i,j}$, where the $\tilde{\epsilon}_{i,{\bf k}}$ are Lagrange
parameters due to the ortho-normality constraint.
In the program a more refined and efficient scheme based on a second order EOM
has been implemented 
\begin{equation}
\label{eq:damped_dynamics}
\frac{d^2}{d\/\tau^2}|\phi_{i,{\bf k}}^{(\tau)}\rangle+2\,\gamma\,
\frac{d}{d\/\tau}| \phi_{i,{\bf
k}}^{(\tau)}\rangle=\left(\tilde{\epsilon}_{i,{\bf k}}-\hat{H}_{\rm KS}\right)
|\phi_{i,{\bf k}}^{(\tau)}\rangle\quad,
\end{equation} where $\gamma$ is a
damping parameter. The EOM is integrated for a step length $\delta\/\tau$ by the
Joannopoulos approach~\cite{payne:86}, which iteratively improves the initial
wave functions. Tassone {\em et~al}.~\cite{tassone:94}
noted that such a scheme is as efficient as the conjugate gradient
techniques~\cite{payne:92}. Though conjugate gradient techniques need a smaller number of
iterations to minimize the energy functional, they need to perform an accurate
line minimization  and have to calculate the
conjugate search directions making this algorithm rather costly compared to the
calculation of the steepest descent direction.  Hence,
the effort to perform an iteration of a conjugate gradient algorithm is
substantially larger
than to perform an iterative step using the damped Joannopoulos algorithm.

In the damped Joannopoulos algorithm the new wave function $|\phi_{i,{\bf
    k}}^{(\tau+1)}\rangle$ is constructed from the wave functions of the last
    two iteration steps $\tau$ and $\tau-1$
\begin{eqnarray}
\label{eq:dmpd_jnpls}
\langle{\bf G}+{\bf k}|\phi_{i,{\bf k}}^{(\tau+1)}\rangle=
\langle{\bf G}+{\bf k}|\phi_{i,{\bf k}}^{(\tau)}\rangle&+&
\beta_{\bf G}\,\langle{\bf G}+{\bf k}|\phi_{i,{\bf k}}^{(\tau)}\rangle\\
&+&\gamma_{\bf G}\,\langle{\bf G}+{\bf k}|\phi_{i,{\bf k}}^{(\tau-1)}\rangle
-\eta_{\bf G}\,\langle{\bf G}+{\bf k}|\hat{H}_{\rm KS}|\phi_{i,{\bf k}}^{(\tau)}\rangle\nonumber
\end{eqnarray}
where the coefficients are
\begin{eqnarray}
\beta_{\bf G}&=&\frac{\tilde{\epsilon}_{i,{\bf k}}\/(h(\delta t)-1)
-\langle{\bf G}+{\bf k}|\hat{H}_{\rm KS}|{\bf G}+{\bf k}\rangle\,{\rm e}^{-\gamma\/\delta t}}
{\tilde{\epsilon}_{i,{\bf k}}-\langle{\bf G}+{\bf k}|\hat{H}_{\rm KS}|{\bf G}+{\bf k}\rangle}\\
\gamma_{\bf G}&=&{\rm e}^{-\gamma\/\delta t}\\
\eta_{\bf G}&=&\frac{h(\delta t)-{\rm e}^{-\gamma\/\delta t}-1}
{\tilde{\epsilon}_{i,{\bf k}}-\langle{\bf G}+{\bf k}|\hat{H}_{\rm KS}|{\bf G}+{\bf k}\rangle}\nonumber
\end{eqnarray}
with the damping parameter $\gamma$, the step length parameter $\delta t$, and 
with \mbox{$\tilde{\epsilon}_{i,{\bf k}}=\langle\phi_{i,{\bf k}}^{(\tau)}|\hat{H}_{\rm KS}|
\phi_{i,{\bf k}}^{(\tau)}\rangle$.} The function $h(\delta t)$ is defined by
\[
h(\delta t)=\left\{
\begin{array}{ll}
{\rm e}^{-\frac{\gamma}{2}\,\delta t}\,{\rm cos}\left(\omega\,\delta\/t\right)
&\,\,\,{\rm if}\,\, \omega^2\geq0\\
{\rm e}^{-\frac{\gamma}{2}\,\delta t}\,{\rm
  cosh}\left(\sqrt{\left|\omega^2\right|}\,\delta\/t\right)&\,\,\,{\rm if}\,\, \omega^2 < 0
\end{array}\right.
\]
with $\omega^2={\tilde{\epsilon}_{i,{\bf k}}-\langle{\bf G}+{\bf
    k}|\hat{H}_{\rm KS}|\phi_{i,{\bf
      k}}^{(\tau)}\rangle}-\frac{\gamma^2}{4}$.

After each iteration step the wave functions $\left\{|\phi_{i,{\bf
    k}}\rangle\right\}$ have to be ortho-normalized, which is done by the Grahm-Schmidt
ortho-normalization scheme. Otherwise all states would converge to the lowest
lying state.

The steepest descent direction which points to the total-energy minimum is given by
\begin{eqnarray}
\langle{\bf G}+{\bf k}|\hat{H}^{\rm KS}|\phi_{i,{\bf k}}\rangle
\label{eq:steepest}
&=&\langle{\bf G}+{\bf k}|-\frac{1}{2}\nabla^2|\phi_{i,{\bf k}}\rangle\\
&&+\langle{\bf G}+{\bf k}|\underbrace{\hat{V}^{\rm H}+\hat{V}^{\rm ps,local}+\hat{V}^{\rm XC}}
_{\hat{V}^{\rm local}}|\phi_{i,{\bf k}}\rangle\nonumber\\
&&+\langle{\bf G}+{\bf k}|\hat{V}^{\rm ps,nl}|\phi_{i,{\bf k}}\rangle\quad.\nonumber
\end{eqnarray} It is calculated very efficiently avoiding expensive vector
matrix products by evaluating\\ \mbox{$\langle{\bf G}+{\bf
  k}|-\frac{1}{2}\nabla^2|\phi_{i,{\bf k}}\rangle$} in the momentum
representation and \mbox{$\langle{\bf G}+{\bf k}|\hat{V}^{\rm local}|\phi_{i,{\bf
    k}}\rangle$} in configuration space representation where
$\frac{1}{2}\nabla^2$ and $\hat{V}^{\rm local}$ are diagonal.  Transformations
between momentum representation and configuration space representation are
performed by fast Fourier transformations and the cost for the calculation of
the latter two expressions is $O(N\ln N)$ operations per state and  k-point~\cite{car:85}.

Though the damped Joannopoulos algorithm is at least twice as efficient as the
first order scheme, given by Eq.~(\ref{eq:steepest_descent}),
additional storage for the wave function
$|\phi^{(\tau-1)}_{i,{\bf k}}\rangle$ is needed. Hence we use the
William-Soler algorithm~\cite{williams:87} whenever storage requirements do
not permit to employ the damped Joannopoulos algorithm.  The coefficients of
this scheme in Eq.~(\ref{eq:dmpd_jnpls}) are
\begin{eqnarray*}
  \beta_{\bf G}&=&\frac{\tilde{\epsilon}_{i,{\bf k}} {\rm
      e}^{\tilde{\epsilon}_{i,{\bf k}}-\langle{\bf G}+{\bf k}|\hat{H}_{\rm
        KS}|{\bf G}+{\bf k}\rangle}} {\tilde{\epsilon}_{i,{\bf k}}-\langle{\bf
      G}+{\bf k}|\hat{H}_{\rm KS}|{\bf G}+{\bf k}\rangle} \\ \eta_{\bf G}&=&
  \frac{{\rm e}^{\tilde{\epsilon}_{i,{\bf k}}-\langle{\bf G}+{\bf k}|
      \hat{H}_{\rm KS}|{\bf G}+{\bf k}\rangle}}
{\tilde{\epsilon}_{i,{\bf k}}-\langle{\bf G}+{\bf k}|\hat{H}_{\rm KS}|{\bf G}+{\bf k}\rangle}
\end{eqnarray*}
with $\gamma_{\bf G}=0$.  The wave functions are improved by successive
 iterations by either of the schemes outlined above until a convergence
 criterion concerning the accuracy of the energy or the forces as described in
 Section~\ref{sec:flwchrt} is fulfilled. On convergence the package proceeds
 with a MD step, a structure optimization step or just terminates.

\paragraph{Mixed-basis-set initialization} 
As pointed out above iterative techniques require an initial guess for the
wave function $|\phi_{i,{\bf k}}^{(0)}\rangle$. Regardless of the method used,
a good initial guess for the initial wave function $|\phi_{i,{\bf
    k}}^{(0)}\rangle$ is essential and can significantly improve the
convergence of the method.  The simplest choice is to generate $|\phi_{i,{\bf
    k}}^{(0)}\rangle$ from random numbers.  However, the trial wave function
obtained by this procedure will be far away from the final solution.
Consequently, less iterations are necessary when the initial wave function is
obtained in a more physical way, i.e. by a direct diagonalization of the
Kohn-Sham Hamiltonian. The effort to diagonalize the Kohn-Sham Hamiltonian on
the full basis, however, outweighs the effort saved in the iterative
diagonalization. Instead, a very efficient procedure is to represent the
Kohn-Sham orbitals $\left\{|\chi_{\nu,{\bf k}}\rangle\right\}$ in a LCAO or
mixed-basis-set which dramatically reduces the number of basis functions and
at the same time gives a good approximation to the solution within the full
basis~\cite{neugebauer:95}. In this basis the Kohn-Sham Hamiltonian is easily
diagonalized for a few number of self-consistency cycles and the overall effort
is considerably reduced.

The mixed basis-set is a subset of the full plane-wave basis. It includes
Bloch-states derived from atomic orbitals and plane waves up to an energy
cut-off $E_{\rm cut}^{\rm init}$ much lower than the energy cut-off of the
full basis.  The localized basis functions are constructed by Grahm-Schmidt
ortho-normalization from the states $|\tilde{\mu}_{I_{\rm s},I_{\rm a},l,m}
\rangle $, which are projections of Bloch-sums onto the plane-wave basis:
\begin{equation}
\label{eq:loc_bas}
|\tilde{\mu}_{I_{\rm s},I_{\rm a},l,m} \rangle = \sum_{\overset{\bf
    G}{\underset{{E_{\rm cut}^{\rm init}\leq\frac{1}{2}|{\bf G}+{\bf k}|^2 \le
        E_{\rm cut}}}{\,}}} \langle {\bf G} + {\bf k} | \mu_{I_{\rm s},I_{\rm
    a},l,m} \rangle \, |{\bf G} + {\bf k}\rangle\quad.
\end{equation}
The functions $\mu_{I_{\rm s},I_{\rm a},l,m}({\bf r})$ are   
\begin{equation}
\mu_{I_{\rm s},I_{\rm a},l,m}({\bf r})  = \sum_{\bf R} {\rm e}^{\immath\,{\bf
    k}\cdot{\bf R}}\,\psi^{\rm ps}_{I_{\rm s},l,m}({\bf r}-{\bf R}-\boldsymbol{\tau}_{I_{\rm
    s},I_{\rm a}})\quad,
\end{equation} where $\psi^{\rm ps}_{I_{\rm s},l,m}({\bf r})$ are the atomic
pseudo wave functions. Using this basis an adequate description of the
localized states as well as of the interstitial region is
possible and the adequate description of strongly
localized states reduces the number of necessary iterations of the iterative
energy minimization scheme~\cite{neugebauer:95}.

In Eq.~(\ref{eq:loc_bas}) the localized basis functions are defined in terms
of a plane-wave representation. Therefore the mixed basis matrix
representation $\langle \chi_{\mu,{\bf k}} | \hat{H}_{KS} | \chi_{\nu,{\bf k}}
\rangle$ of the Kohn-Sham operator can be easily calculated using the same
routines as employed to compute the steepest descent vector in
Eq.~(\ref{eq:steepest}) just at the additional cost of a scalar product:
\begin{equation}
\langle \chi_{\mu,{\bf k}} | \hat{H}_{KS} | \chi_{\nu,{\bf k}} \rangle =
\sum_{\bf G} \langle  \chi_{\mu,{\bf k}} | {\bf G} + {\bf k} \rangle
\langle {\bf G} + {\bf k} | \hat{H}_{KS} | \chi_{\nu,{\bf k}} \rangle\quad.  
\end{equation} This fact combined with the small size of the basis reduces the
computational effort of the matrix diagonalization approximately to the
cost of an energy minimization step within the full basis.

This procedure is repeated until electronic self-consistency is reached. Old
and new electron densities are mixed using the Broyden-scheme~\cite{singh:86}.
The resulting Kohn-Sham orbitals are used as initial wave functions for the
iterative energy minimization scheme described above

\paragraph{Implementation}
The program {\sf fhi96md} efficiently performs on low budget personal
computers even for large systems. Obviously, it works even better on high
performance computers. The high efficiency of the code is achieved by a low
storage demand and an economic cache utilization. For example single precision
arrays are used to {\em store} the wave function coefficients $c_{i,{\bf G}+{\bf
k}}$ and the reciprocal lattice vectors ${\bf G}$ without effecting the
accuracy of the results. The portability of the code to a variety of platforms
without loosing efficiency is achieved by using standard library subroutines
like BLAS-routines. In this way on each platform full advantage can be taken
of optimized, platform specific implementations of the libraries.  However,
when such libraries are not available (e.g. on a PC) we provide the necessary
routines in a separate library which can be linked to the program.

In typical applications the calculation of the electron density and the
steepest descent direction dominate the computational effort (mainly
FFT). These contributions scale like \mbox{$O(N^2\,{\rm ln}\/N)$} with the
system size. In very large systems, e.g. a Silicon super cell with 128 atoms
and more, the calculation of the (non-local contributions of the) atomic
forces (which scale like $O(N^3)$ becomes the dominant contribution. The
computational effort for this contribution is dramatically reduced by
calculating the forces only for the last few electronic steps, when the
Born-Oppenheimer surface is almost reached.
\subsection{Integrating the equations of motion of the nuclei}
\label{sec:dyn}
Once the ground state of the electrons is calculated as described in
Sec.~\ref{sec:elproblem} the atomic EOMs are integrated with
standard MD techniques.  We have implemented two schemes, the Verlet
algorithm and a predictor-corrector algorithm. The choice of the scheme
depends on the actual type of application.  For example, the Verlet algorithm
is more stable with respect to an energy drift than the predictor-corrector
algorithm. However, when it is important to obtain very accurate velocities,
the predictor-corrector algorithm should be employed.

The Verlet algorithm~\cite{verlet:67} is
\begin{equation}
\boldsymbol{\tau}_{I_{\rm s},I_{\rm a}}(t+\delta\/t_{\rm nuc})=2\/\boldsymbol{\tau}_{I_{\rm s},I_{\rm a}}(t)
-\boldsymbol{\tau}_{I_{\rm s},I_{\rm a}}(t-\delta\/t_{\rm nuc})+\delta\/t_{\rm nuc}^2\,\frac{{\bf
    F}_{I_{\rm s},I_{\rm a}}(\{\boldsymbol{\tau}_{I_{\rm s},I_{\rm a}}(t)\})}{M_{I_{\rm s}}} 
\end{equation} where $M_{I_{\rm s}}$ is the mass of the nuclei and
$\delta\/t_{\rm nuc}$ is the time step. From a numerical point of view it is
desirable to chose a time step $\delta\/t_{\rm nuc}$ as large as possible. A
good choice for $\delta\/t_{\rm nuc}$ is $\frac{1}{15}$ of the shortest phonon
period in the system, which is approximately {\em one} order of magnitude
larger than the time step used in the Car-Parrinello {\em ab initio}
MD~\cite{car:85}. The time step there is limited by the fast oscillatory
motion of the electrons in the fictitious electron dynamics~\cite{pastore:91},
which originates from the need to keep the electrons close to the
Born-Oppenheimer surface.


The velocity ${\bf v}_{I_{\rm
    s},I_{\rm a}}$ of a nucleus is calculated by
\begin{equation}
\label{eq:v_verlet}
{\bf v}_{I_{\rm s},I_{\rm a}}(t)=\frac{1}{2\,\delta\/t_{\rm nuc}}
\left(3\/\boldsymbol{\tau}_{I_{\rm s},I_{\rm a}}(t)-4\/\boldsymbol{\tau}_{I_{\rm s},I_{\rm
    a}}(t-\delta\/t_{\rm nuc})+\boldsymbol{\tau}_{I_{\rm s},I_{\rm a}}(t-2\,\delta\/t_{\rm
  nuc})\right)\quad,
\end{equation} which is correct to second order in $\delta\/t_{\rm
  nuc}$~\cite{bloechl:92}. Note that the velocities are not directly
integrated by the algorithm, but obtained from the trajectory itself.

In a predictor-corrector scheme the discretization error is of higher order in
$\delta\/t_{\rm nuc}$ than that of the Verlet algorithm. The usual
predictor-corrector (PC) schemes employ an extrapolation step to predict
positions and velocities, which are then corrected in the corrector step. For
small time steps the accuracy of the PC scheme is much higher than that of the
Verlet algorithm. However, for a large time step it looses this
advantage~\cite{allen:87}, since the error in the extrapolation step increases
strongly with an increasing time step.  Therefore we have implemented a
predictor-corrector scheme which avoids the extrapolation. In this scheme an
Adams-Bashforth predictor step
\begin{eqnarray}
\boldsymbol{\tau}^0_{I_{\rm s},I_{\rm a}}(t_n)&=&\boldsymbol{\tau}_{I_{\rm s},i_{\rm
    a}}(t_{n-1})+\delta\/t_{\rm nuc}\,\sum_{k=1}^q
\alpha_{q,k} {\bf v}_{I_{\rm s},I_{\rm a}}(t_{n-k})\nonumber\\
{\rm and}\quad\quad\quad\quad\quad\quad\quad&&\\
{\bf v}^0_{I_{\rm s},I_{\rm a}}(t_n)&=&{\bf v}_{I_{\rm s},i_{\rm
    a}}(t_{n-1})+\delta\/t_{\rm nuc}\,\sum_{k=1}^q
\alpha_{q,k} \frac{{\bf F}_{I_{\rm s},I_{\rm a}}(\{\boldsymbol{\tau}_{I_{\rm s},I_{\rm a}}(t_{n-k})\})}{m_{i_{\rm
      s}}}\nonumber
\end{eqnarray}
is followed by an Adams-Moulton corrector step
\begin{eqnarray}
 \boldsymbol{\tau}_{I_{\rm s},I_{\rm a}}(t_n)&=&\boldsymbol{\tau}_{I_{\rm s},i_{\rm
    a}}(t_{n-1})+\delta\/t_{\rm nuc}\,\beta_{q,0}\, {\bf v}^0_{I_{\rm s},i_{\rm
    a}}(t_{n})
+\delta\/t_{\rm nuc}\,\sum_{k=1}^q\beta_{q,k} {\bf v}_{I_{\rm s},i_{\rm
    a}}(t_{n-k})\nonumber\\ 
{\rm and}\quad\quad\quad\quad\quad\quad\quad&&\\
 {\bf v}_{I_{\rm s},I_{\rm a}}(t_n)&=&{\bf v}_{I_{\rm s},i_{\rm
    a}}(t_{n-1})+\delta\/t_{\rm nuc}\,\beta_{q,0}\, \frac{{\bf F}_{I_{\rm s},i_{\rm
    a}}(\{\boldsymbol{\tau}^0_{I_{\rm s},I_{\rm a}}(t_n)\})}{m_{I_{\rm s}}}\nonumber\\
&&\quad\quad\quad\quad\quad\quad+\delta\/t_{\rm nuc}\,\sum_{k=1}^q\beta_{q,k} \frac{{\bf F}_{I_{\rm s},i_{\rm
    a}}(\{\boldsymbol{\tau}_{I_{\rm s},I_{\rm a}}(t_{n-k})\})}{m_{I_{\rm s}}}\nonumber
\end{eqnarray} with the coefficients $\alpha_{q,k}$ and $\beta_{q,k}$ as
tabulated in Ref.~\cite[pp 126]{stoer:78} and with $t_{n}=t+n\,\delta\/t_{\rm
  nuc}$.  Though in this scheme the forces have to be re-calculated after the
corrector step, this is inexpensively done as the corrector step corrects the
positions of the nuclei only slightly and convergent forces are typically
obtained within a few iterative steps. Thus the accuracy of the
predictor-corrector scheme does not suffer from a large time step. Moreover,
the velocities are smooth functions of time and the total-energy of the
coupled electron nuclei system is practically free from fluctuations.

Energy losses encountered in long simulations within the micro-canonical
ensemble can be compensated by a periodic rescaling of the atomic velocities
according to a preset average kinetic energy per particle.  Temperature is
optionally enabled by a Nos\'e-Hoover thermostat~\cite{nose:84,hoover:85}.
One additional degree of freedom is employed to simulate the heat bath and an
extra force is accelerating or de-accelerating the nuclei to maintain the
temperature of the system. The modified EOMs of the extended system are
\begin{eqnarray} 
 &&M_{\rm is}\,\frac{d^2}{d\/t^2}\boldsymbol{\tau}_{I_{\rm s},I_{\rm a}}={\bf F}_{I_{\rm
     s},I_{\rm a}} - \frac{d}{d\/t}s\,{\bf v}_{I_{\rm s},I_{\rm a}}
\nonumber\\
{\rm and}\,\,\,\,\,\,\,\,\,\,\,&&\\
&&Q\,\frac{d^2}{d\/t^2}s=\sum_{I_{\rm s},I_{\rm a}}M_{I_{\rm s}}\/{\bf
  v}^2_{I_{\rm s},I_{\rm a}}-g\,k_{\rm B}T\nonumber
\end{eqnarray}
where $g$ is the number of independent degrees of freedom in
the system. 

The initial coordinates and velocities of all independent degrees of freedom
completely determine the trajectory. The initial velocities are generated
from Gaussian distributed random numbers such that the center of mass is at
rest and that the kinetic energy of the system corresponds to an initial
temperature. In a micro-canonical ensemble the initial temperature together
with the potential energy of the initial configuration of the nuclei determine
the resulting average kinetic energy per particle. 

Both schemes, the Verlet algorithm and the predictor-corrector scheme, require
knowledge of the trajectory at previous time steps. Therefore, to start these
algorithms, the Runge-Kutta scheme~\cite{stoer:78} is used to integrate the
equation of motion for the first few time steps.
\subsection{Structure optimization}
\label{sec:relax}
An efficient and numerically stable method to find the equilibrium geometry is
damped Newton dynamics. Starting from an initial configuration the nuclei are
moved according to the iterative scheme
\begin{equation}
\boldsymbol{\tau}_{I_{\rm s},I_{\rm a}}^{(n_{\rm it}+1)}=(1+\lambda_{I_{\rm s}})\/\boldsymbol{\tau}_{I_{\rm s},I_{\rm a}}^{(n_{\rm it})}
-\/\lambda_{I_{\rm s}}\,\boldsymbol{\tau}_{I_{\rm s},I_{\rm a}}^{(n_{\rm it}-1)}+\/\mu_{I_{\rm s}}\,
{\bf F}_{I_{\rm s},I_{\rm a}}\left(\left\{\boldsymbol{\tau}_{I_{\rm s},I_{\rm a}}^{(n_{\rm
    it})}\right\}\right), 
\end{equation} where $\lambda_{I_{\rm s}}$ and $\mu_{I_{\rm s}}$ are the
damping and reciprocal mass parameters rsp. The parameters $\lambda_{i_{\rm
    s}}$ and $\mu_{I_{\rm s}}$ determine whether the nuclei loose their
    initial potential energy slowly in an oscillatory-like motion or whether
    they move straight into the closest local minimum. The program allows also
    to confine the configuration space open to the search. An example is the
    calculation of adiabatic potential energy surfaces, where the ad-atom or a
    defect is held fixed and all other atoms are allowed to relax.
\section{The package fhi96md}
\label{sec:impl}
The package contains the program {\sf fhi96md} and a start utility {\sf
  start}. The program {\sf fhi96md} performs the MD simulations and the
total-energy calculations. The start utility generates the file {\file
  parameter.h} containing the parameters and the file {\file input.ini}
containing the input data, which are necessary to compile and to run the
program {\sf fhi96md}.  Features of the start utility include an automatic search
for point group symmetries and the symmetry center in the system. Further, it
automatically optimizes the parameter file to minimize the memory demand for
each individual run.

The following two Subsections describe the structure of the
program {\sf fhi96md} and discuss the input files processed by the
start utility and the output files generated by {\sf fhi96md}.

\subsection{The program structure}
\label{sec:flwchrt}
The structure of the program {\sf fhi96md} is sketched in
Fig.~\ref{fig:flwchrt_main}. The self-consistent calculation of the electron
ground state forms the main body of the program, which is displayed on the
left-hand side of Fig.~\ref{fig:flwchrt_main}.
The movement of the atoms is
accomplished in the block ``move atoms'', which is sketched on the right hand side
of Fig.~\ref{fig:flwchrt_main}.  Note, that the generation of output is not
explicitly accounted for in the flowchart and we refer to it at the end of
this Subsection.

The first block in the flowchart is the initialization block, where the
program reads the input files {\file inp.mod}, {\file inp.ini} and the
pseudopotential data. Then the routines calculating form factors, structure
factors and phase factors (c.f. Appendix~\ref{sec:app_form_struct}) are called
and the initial wave functions are set up either from a restart file or by a
few self-consistency cycles using the mixed basis-set initialization
(c.f. Section~\ref{sec:elproblem}).  Having obtained the initial wave function
$|\Psi_{i,{\bf k}}^{(0)}\rangle$ the program enters the self-consistency loop.
First, the electron density and the contributions to energy, potential and
forces are calculated.  Note, that the forces are only calculated during MD
simulations and structure optimization when the electrons are sufficiently
close to the Born-Oppenheimer surface.

Within the block ``move atoms'' the atomic EOMs are integrated for one time
step in a MD simulation or a structure optimization is performed, provided the
electrons are sufficiently close to the Born-Oppenheimer surface, i.e. the
forces are converged. The control over the calculation of the forces is
handled by this block. If the nuclei have been moved, i.e. either the atomic
EOMs have been integrated for another time step or one structure optimization
step has been executed, it also recalculates the structure and phase factors
and other quantities that explicitly depend on the positions of the nuclei.
Upon the first call to the routine {\tt fiondyn} in this block the restart
file {\file fort.20} is read, if provided, and all necessary steps are taken
to restart or initialize the dynamics.

The following two blocks update the wave functions using the damped
Joannopoulos or the William-Soler algorithm and ortho-normalize the wave
function by a Grahm-Schmidt ortho-normalization.  In the last block within the
self-consistency loop the occupation numbers are updated e.g. for a metallic
system according to a Fermi distribution by a damped pseudo-eigenvalue
scheme~\cite{stumpf:94,pederson:91}.  This block enables also an interactive
control over the step length $\delta\/\tau$ and damping parameter $\gamma$ of
the energy-minimization scheme, the mass parameter $\mu_{I_{\rm s}}$ of the
structure optimization scheme and the remaining numbers of iterations while
the program is running.  These parameters are updated from the files {\file
delt} ($\delta\/\tau$), {\file ion\_fac} ($\mu_{I_{\rm s}}$), {\file stopfile}
(remaining number of electronic iterations) and {\file stopprogram} (remaining
number of structure optimization steps).  If these files are empty the
parameters are not changed.  Finally, the convergence criteria are
checked. The program terminates when convergence is achieved or when the
preset number of iterations or the allowed CPU-time is exceeded.

Output is generated at the last block of the self-consistency loop and by the
routines {\tt fiondyn}, {\tt fionsc}, {\tt fermi} and {\tt vofrho}. The
routines {\tt fiondyn} and {\tt o\_wave}  generate restart files for MD
simulations and total-energy calculations.

In the mixed-basis-set initialization, the self-consistency loop closely
follows the organization of that discussed above. First of all the initial
electron density is obtained either from a superposition of contracted atomic
pseudo densities or from an electron density of a previous calculation (file
{\file fort.72}). The local contributions to the potential and the energy are
calculated by the routine {\tt vofrho}. The localized orbitals to construct
the mixed-basis-set are set up by routine {\tt project\_init}.  The
non-local contributions to the potential and the energy in the localized basis
set are calculated by the routine {\tt nlrhkb\_b0}.  In the second step the
Hamiltonian is constructed with the help of routine {\tt dforce\_b0}.  The
Hamiltonian is diagonalized by standard diagonalization routines.  The new
electron density is calculated (routine {\tt rho\_psi}).  Finally the new
electron density is mixed with the old density by a Broyden mixing (routine {\tt
broyden}).
\subsection{Input and output files}
Two input files are required as input for the start utility. The file {\file
  inp.mod} contains the control parameters for the run. The file {\file
  start.inp} describes the geometry of the super cell, the configuration of
the nuclei and parameters relevant for the MD
simulation or the structure optimization, and the calculation of the electron
ground state.  The parameters and data contained in the files {\file inp.mod}
and {\file start.inp} are described in Tables \ref{tab:inp_mod_1} and
\ref{tab:start_inp_1}.  
Related entries are grouped in one line as shown in the
listings of the input files of the test run in Section TEST RUN.  No specific
format is requested other than that implied by the type of the entries as
displayed in the second column of Tables \ref{tab:inp_mod_1} and
\ref{tab:start_inp_1}.  The start utility processes the files {\file inp.mod}
and {\file start.inp} and generates the parameter file {\file parameter.h} and
the input file {\file inp.ini}.  These files are described in Appendix~\ref{sec:app_tables}.

In the following we describe the strategy to set up the files {\file
inp.mod} and {\file start.inp}. First of all there are the
logical parameters {\it tdyn} and {\it tfor} - we refer to parameters in the
file {\file inp.mod}, if not noted otherwise. With these two parameters one
instructs {\bf fhi96md} to perform a MD simulation, a structure optimization,
or if both are set to {\tt .false.} just to calculate the electronic structure
for the given configuration of the nuclei. 

For a MD simulation the parameters {\it i\_dyn}, {\it norder}, and {\it
  delt\_ion} specify the scheme for integrating the equation of motion of the
nuclei and the time step (c.f. Section~\ref{sec:dyn}). Additional parameters in
the file {\file start.inp} determine the simulation ensemble ({\it nthm}, {\it
  Q}, {\it T\_ion} and {\it nfi\_rescale} ), the set up of initial positions
and velocities ({\it npos} and {\it coordwave} -- restart options included)
and the masses of the nuclei $M_{I_{\rm s}}$. The parameters needed for the structure
optimization $\mu_{I_{\rm s}}$ and $\lambda_{I_{\rm s}}$ ({\it ion\_fac} and
{\it ion\_damp}) are specified in {\file start.inp} as well (c.f. Section~\ref{sec:relax}).

The geometry of the super cell, the positions of the nuclei and optionally the
velocities (c.f. {\it npos}) are specified in the file {\file start.inp}. The
parameter {\it ibrav} and {\it pgind} determine the lattice type and the point
group symmetry of the configuration of the nuclei. With $pgind=0$ an automatic
search for the point group symmetries and the symmetry center is performed. For
$pgind>1$, the symmetry center is the origin $(0,0,0)$ of the super cell. In MD
simulations and structure optimization point group symmetries are usually not
applicable. For each of the {\it nsp} atomic species one declares the
properties of the pseudopotential ({\it zv}, {\it l\_max} and {\it l\_loc})
and the radius of the screening charge $n^{\rm Gau\ss}_{i_{\rm s}}({\bf r})$
({\it rgauss}, c.f.  Appendix~\ref{sec:expr}). The positions of the nuclei
{\it tau0} and optionally also the velocities {\it vau0} follow this
declaration (c.f. {\it npos}).
Note, that lines containing the velocity of a nuclei immediately follow the line with
the corresponding coordinates. 
Pseudopotential data needs to be provided in the files {\file fort.11}, {\file
fort.12},
\ldots for the {\it nsp} atomic species. The data in each file is expected either in a parameterized form or
tabulated on a logarithmic mesh (parameter {\it tpsmesh}).

Having set up the basic configuration data, we now turn to the data needed for
the calculation of the electron ground state. The parameter {\it ecut}
specifies the energy cut-off of the plane-wave basis-set. The data {\it xk}
and {\it wkpt} declare the {\it nkpt}  k-points ({\it t\_kpoint\_rel} specifies
the frame of reference). The start utility reduces the  k-point set according to the
point group symmetries as specified by the parameter {\it pgind}. 
A special k-point set according to the Monkhorst-Pack
scheme~\cite{monkhorst:76} is generated using the k-point
$(\frac{1}{2},\frac{1}{2},\frac{1}{2})$ with the weight $w_{\bf k}=1.0$ ({\it
t\_kpoint\_rel}={\tt .true.}). The number of mesh points of the k-point mesh
spanned in the Brillouin-zone is specified by {\it i\_fac}. The
k-point set is then given by the irreducible part of the k-point mesh
(c.f. the example in Section TEST RUN).

When the
initial wave functions $|\Psi^{(0)}_{i,{\bf k}}\rangle$ are calculated by an
explicit diagonalization of the Kohn-Sham operator in a mixed-basis-set
(parameter {\it nbeg}) further specifications of the mixed-basis-set and of the
set up of the initial electron density ({\it init\_basis} and {\it nrho}) are
needed. The parameters {\it ecuti} and {\it t\_init\_basis} in the file {\file
  start.inp} specify the energy cut-off $E_{\rm cut}^{\rm init}$ and the
atomic pseudo orbitals included in the mixed-basis-set. The number of
self-consistency cycles of the initialization is declared by {\it
  nomore\_init}.

The approximation to the exchange-correlation energy functional is
determined by the parameter {\it i\_xc}. For applying the GGA correction only
in the last step of an LDA calculation set the parameter {\it tpostc} to {\tt
.true.}. The program applies non-linear core valence
exchange and correlation automatically, when pseudopotentials
that have been generated with a pseudo core density are supplied.

For surface calculations additional
contributions to the energy due to a  correction for a surface dipole moment may
be included (parameter {\it tdipol}). This requires a surface in the xy-plane~\cite{neugebauer:92}.

The scheme used in the iterative diagonalization is specified by {\it
  i\_edyn}. The corresponding parameters $\delta\/t$ and $\gamma$ of the
damped Joannopoulos and William-Soler algorithm are {\it delt} and {\it
  gamma}. Note, that when the total-energy improvement per iteration is less
than {\it eps\_chg\_delt} the parameters {\it delt2} and {\it gamma2} are used
instead, accelerating the convergence of the scheme (with {\it delt}$>${\it
  delt2}). The choice for {\it delt} and {\it gamma} strongly depends on the
atomic species and configuration. However, a good guess for $\delta\/t$ lies
in the range $1.0< delt <20.0$ and a good choice for $\gamma$ is $\gamma\sim
0.2$. 

In MD simulations and structure optimization the routine {\tt fiondyn} and
{\tt fionsc} monitor the error of the forces ({\it force\_eps}) before
integrating the equations of motion of the nuclei or performing a structure
optimization step. Note, however, that the forces are not calculated for a
specified number of iterations ({\it max\_no\_force}) and unless the electrons
are sufficiently close to their ground state ({\it epsel} and {\it
epsekinc}). The structure optimization terminates, if the residual forces
acting on the nuclei are sufficiently small ({\it epsfor}).  The calculation of
the electron ground state for a fixed configuration of the nuclei is
terminated, if the improvement of the total-energy and the wave functions per
iteration is smaller than {\it epsel} and {\it epsekinc} rsp.  Nevertheless
{\sf fhi96md} is terminated on exceeding either a maximum number of steps
({\it nomore} and {\it nstepe}) or the overall CPU-time limit -- which is
of importance when the program is running in queuing-system imposing CPU-time
restrictions.

A proper setting of the parameters {\it pfft\_store} and {\it mesh\_accuracy}
provides an even better performance of the code. The parameter {\it
  pfft\_store} determines the fraction of wave functions of which the real
space representation is stored to avoid a second transformation. The only
limitation is the available physical memory. The parameter {\it mesh\_accuracy}
specifies the fraction of Fourier coefficients used to represent the
electron density (c.f. Appendix~\ref{sec:app_tables}).  Choosing {\it
  mesh\_accuracy}$\,=1.0$ implies a proper treatment of the electron density
without approximations. In many systems, a value of $0.8$ results in acceptable loss in
accuracy and at the same time in a much better performance.

Output generated during the calculation is written to several files. The chief
output file is {\file fort.6}. It contains a complete protocol of the
initialization, the molecular dynamic simulation or the structure optimization
and information on each step of the energy minimization. During the molecular
dynamics simulation the trajectory is also written to the unformatted file
{\file fort.2} (c.f. routine {\tt fiondyn}). The file {\file fort.1} contains
information on the position of the nuclei and forces, written at each
self-consistency cycle, while performing structure optimizations or MD
simulations.

Restart files are written every {\it iprint} self-consistency cycles (c.f.
{\file inp.mod}). The file {\file fort.71} contains among others the wave
functions and the coordinates of the nuclei. A restarting run reads this file
renamed {\file fort.70}.  The electron density is stored in file {\file
  fort.72}. The file {\file fort.21} contains all necessary restart
information of the molecular dynamics.
\section{Making of the program}
The package {\sf fhi96md} is available as a tar-archive and can be extracted by
the UNIX command {\em tar}. The directory {\sf fhi96md} is the root of the
package's directory tree.  The directory {\sf bin} contains shell scripts
for the test run and other examples.  These shell-scripts create the
input files {\file inp.mod} and {\file start.inp}, compile and run the start
utility to generate the input and parameter files for {\sf fhi96md} and
finally compile and run the program {\file fhi96md}. Pseudopotentials for the
example runs are included in the directory {\file pseudo}. They have been
generated according to the schemes of Hamann~\cite{hamann:89}. Directories
with the generic name {\file example.}$<scriptname>$ contain the formatted
output of the examples and the test run.

The directory {\file src} houses all sources and libraries of the
package.
The sources of the utility {\sf start} and the program {\sf
fhi96md} are stored in the corresponding subdirectories. Also included in these directories
are the makefiles used in the examples to compile the programs by the UNIX command {\em make}.

Libraries are contained in the subdirectory {\file lib} together with makefiles and sources. 
These libraries are automatically compiled as recommended by the makefiles of the start 
utility and the program {\file fhi96md}. However, there are still some library routines like the
fast Fourier transformation, EISPACK-routines and BLAS-routines which we link from commercial libraries such as 
the ESSL. These routines are adopted to the platform running the program and 
utilize a higher performance than public domain routines would do. Currently we offer
three versions compatible with the ESSL-library, the IMSL-library and the public domain BLAS
and FFTPACK routines as contained in the netlib.

In order to run the program on other platforms, first the makefiles have to be
adopted, i.e. the FORTRAN77 compiler name as well as the compiler and linker
options have to be set properly.  Automatic zero-initialization of all
variables is recommended.

Second, the available library needs to be adjusted in all example shell scripts
as contained in the directory {\file bin}, i.e. one needs to replace the
option {\tt essl} at the make command call by either {\tt imsl} or {\tt
netlib}. Two routines in {\file libnum} have to be ported to
the specific platform: the routine {\file cputime} -- measuring the elapsed
CPU-time between subsequent calls -- and the routine {\file flush} -- flushing the file buffer.

Note, that some large arrays are declared single precision ({\tt real*4} and
{\tt complex*8}) to reduce the memory demand. This may require some
adjustment in the calls of precision depending routines of the external
library when single and double precision are used with another convention as
e.g. on CRAY supercomputers.
\section{Test run} The test run simply calculates the electron ground state of
a bulk GaAs cell. The lattice is a simple cubic bravais lattice and the super
cell contains eight gallium and arsenic atoms. The calculation is performed
with an energy cut-off of $8$ Ry and 4 special k-points in the
irreducible wedge of the Brillouin-zone. The initial diagonalization is
performed in a basis-set containing plane waves up to an energy cut-off of $4$
Ry.

The shell-script {\file run.GaAs.bulk} performing the test run creates the input files 
{\file inp.mod} and {\file start.inp}, compiles and runs the start utility and finally 
compiles and runs the program {\file fhi96md}. The workspace is the directory {\file work},
where all output of either {\file start} and {\file fhi96md} is stored. The pseudopotential 
files {\file ga\_aa.cpi} for gallium and {\file as\_aa.cpi} for arsenic are provided in the directory {\file pseudo}
and are copied by {\file run.GaAs.bulk}  to the workspace. The pseudopotentials~\cite{fuchs:96} have been created according to the scheme of Hamann~\cite{hamann:89}. The 
input files {\file inp.mod} and {\file start.inp}, the files {\file parameter.h} and 
{\file inp.ini}, and extracts from the major output file
{\file fort.6} are found at the end of the paper. The full set of formatted output files
is contained in the directory {\file example.GaAs.bulk}.
\clearpage
\appendix
\section{List of expressions}
\label{sec:expr}
In this Appendix we list the central physical quantities and expressions as
implemented in the routines of the package {\sf fhi96md}. It is organized in
three Subsections: for routines calculating charge densities, for routines
predominantly dealing with electronic contributions to energy, potentials and
forces and for routines tabulating structure factors and form factors.  The
sequence in which these routines are called is described in
Section~\ref{sec:flwchrt} and in the flow chart in
Fig.~\ref{fig:flwchrt_main}.

Throughout this Appendix and in the package atomic units are used -- energies
are given  in units of
Hartree -- unless noted otherwise and with the exception that in the package reciprocal lattice vectors are in units of 
${2\/\pi}/{a_{\rm lat}}$. The angular momentum quantum number $l$ translates into 
an index equal to $l+1$ in any routine.
The weights $w_{\bf k}$ of k-points are as
  given in the input files, though in the package the $w_{\bf k}$ 
are eventually divided
  by the number of point group elements {\it nrot}. Hence an additional factor
  {\it nrot} may appear. 
The symbols in this Appendix are as defined in the text below and the variable 
name is stated whenever this is of importance. Some 
general symbols are listed in Table~\ref{tab:def}. Be aware that
a few variables in the package may store different quantities (e.g. in {\it rhoe}
the electron density and the local potential are stored).
For the sake of compact formulas, summation symbols only bear the summation
index, the associated range of summation is listed in Table~\ref{tab:range} .
\subsection{Charge density of valence electrons and pseudo core} 
\paragraph{Routine {\tt rhoofr}}calculates the electron density and the
kinetic energy:

The wave function is given by $\Psi_{i,{\bf k}}({\bf r})={\rm e}^{\immath {\bf
k}\cdot{\bf r}}\,u_{i,{\bf k}}({\bf r})$ with:
\begin{equation}
u_{i,{\bf k}}({\bf r})= \sum_{\bf G}\,{\rm e}^{\immath {\bf G}\cdot{\bf
r}}\,c_{i,{\bf G}+{\bf k}}\quad{\rm
and}\quad\int_{\Omega}d^3\/r\,\overline{u_{i,{\bf k}}({\bf r})}\/u_{j,{\bf
k}}({\bf r}) =\Omega\,\delta_{i,j},
\end{equation}
where the ortho-normalization is accomplished by routine {\tt graham}.\\
The electron density $n({\bf r})$ -- variable {\it rhoe} -- is given by
\begin{equation}
n({\bf r})=\frac{1}{\Omega}\sum_{s}\sum_{{\bf k}}\sum_{i}\,
w_{\bf k}\/f_{i,{\bf k}} |u_{i,{\bf k}}(s^{-1}{\bf r})|^2\quad.
\end{equation}
The symmetrization is performed as the last step in real space.\\
The kinetic energy -- variable {\it ekin} -- is obtained from:
\begin{equation}
T^{\rm s}=\frac{1}{2}
\sum_{\bf k}\sum_{i}\/w_{\bf k}\/f_{i,{\bf k}}
\sum_{\bf G} |{\bf G}+{\bf k}|^2 |c_{i,{\bf G}+{\bf k}}|^2
\label{eq:ekin}
\end{equation}
\paragraph{Routine {\tt corcha}}calculates pseudo core charge density
and tabulates the form factors:

The pseudocore density $\tilde{n}^{\rm core}({\bf r})$ is calculated by:
\begin{equation}
\label{eq:core_chrg}
\tilde{n}^{\rm core}({\bf r})=\sum_{\tilde{\bf G}}\/{\rm e}^{\immath\,{\tilde{\bf G}}\cdot{\bf r}}\/
\sum_{I_{\rm s}}\/S_{I_{\rm s}}({\tilde{\bf G}})\/\Phi_{I_{\rm s}}^{\rm core}({\tilde{\bf G}})\quad,
\end{equation}
where $\tilde{n}^{\rm core}_{I_{\rm s}}({\bf r})$ is the form factor of
pseudocore density -- variable {\it formf\_at}:
\begin{equation}
\Phi^{core}_{I_{\rm s}}({\bf G})=\frac{4\/\pi}{\Omega}\,\int_{0}^{\infty}d\/r\,r^2
\,j_0(|{\bf G}|\/r)\tilde{n}^{core}_{I_{\rm s}}(r)\quad.
\end{equation}
The sum in Eq.~(\ref{eq:core_chrg}) includes only species for which a pseudo core
$\tilde{n}^{\rm core}_{I_{\rm s}}({\bf r})$ is included in the creation of the
pseudopotential. 
\subsection{Energy, potential and forces: electronic contribution}
\label{sec:app_expr_el_contrib}
\paragraph{Routine {\tt vofrho}}calculates the local contributions to energy,
potential and forces:\\ 

Due to the long-range tail of the coulomb potential $V^{\rm H}({\bf G})$ and
$V^{\rm ps,local}({\bf G})$ diverge in a periodic system, though in the sum of
the two potentials the divergent terms cancel. In order to treat the
potentials separately a charge density $n^{\rm Gau\ss}({\bf r})$ is introduced
to remove the divergent contributions without affecting the sum of the
potentials.  The energy contributions $E^{\rm H}$, $E^{\rm ps, local}$ and
$E^{\rm nuc-nuc}$ are treated accordingly. By means of $n^{\rm Gau\ss}({\bf
r})$ these terms are redefined
\begin{equation}
E^{\rm H}[n]+E^{\rm ps,local}+E^{\rm nuc-nuc}\rightarrow E^{\rm H}[n+n^{\rm
Gau\ss}]+\tilde{E}^{\rm ps,local}+E^{\rm sr}-E^{\rm self}\quad.
\end{equation}
For the explicit expressions see Eqn. (\ref{eq:potential}), (\ref{eq:gauss})
and (\ref{eq:etot}).\\

The contributions to the local potential in reciprocal space are:
\begin{equation}
V^{\rm local}({\bf G})=V^{\rm H}({\bf G})+V^{\rm ps,local}({\bf G})+V^{\rm XC}({\bf G})\quad.
\end{equation}
In surface calculations an additional contribution $V^{\rm dipole}({\bf r})$ to the local 
potential $V^{\rm local}({\bf r})$ arises according to Eq.~(\ref{eq:vdipole}) due to a surface
dipole moment.

The Hartree potential is obtained from:
\begin{equation}
\label{eq:potential}
V^{\rm H}({\bf G})=\frac{4\/\pi}{|{\bf G}|^2} \left(n({\bf G})+n^{\rm Gau\ss}({\bf G})\right)\quad,
\end{equation}
with
\begin{equation}
n({\bf G})=\frac{1}{\Omega}\/\int_{\Omega}\/d^3r\,{\rm e}^{-\immath\/{\bf G}\cdot{\bf r}}\,n({\bf r})
\end{equation}
and
\begin{equation}
\label{eq:gauss}
n^{\rm Gau\ss}({\bf G})=\sum_{I_{\rm s}}\,S_{I_{\rm s}}({\bf G})\/
\Phi^{\rm Gau\ss}_{I_{\rm s}}({\bf G})\quad,
\end{equation}
where $\Phi^{\rm Gau\ss}_{I_{\rm s}}({\bf G})$ is given by Eq.~(\ref{eq:rhops}).\\
The local pseudopotential $V^{\rm ps,local}({\bf G})$ reads:
\begin{equation}
V^{\rm ps,local}({\bf G})=\sum_{I_{\rm s}}\,S_{I_{\rm s}}({\bf G})\/
\Phi^{\rm ps}_{I_{\rm s}}({\bf G})
\end{equation}
with $S_{I_{\rm s}}({\bf G})$ and $\Phi^{\rm ps}_{I_{\rm s}}({\bf G})$ 
given by Eqn. (\ref{eq:struct}) and (\ref{eq:vps}) respectively.\\
The exchange-correlation potential is calculated from:
\begin{equation}
V^{\rm XC}({\bf G})= \frac{1}{\Omega}\/\int_{\Omega}\/d^3r\,{\rm e}^{-\immath\/{\bf G}\cdot {\bf r}}
\,V^{\rm XC}\left(n({\bf r})+\tilde{n}^{\rm core}({\bf r})\right)\quad,
\label{eq:vxc}
\end{equation}
where the pseudocore density is included, if provided with the pseudopotentials. 
$V^{\rm XC}[n]$ is approximated by LDA, BP, or PW91 functional and in the latter 
two cases the first and second derivatives of the electron and pseudocore density are evaluated 
in routine {\it stvxc\_gc}:
\begin{eqnarray}
\nabla\/n({\bf r})&=& \sum_{\tilde{\bf G}} \immath\/{\tilde{\bf G}}\,{\rm e}^{\immath\/{\tilde{\bf G}}\cdot {\bf r}}\,n({\tilde{\bf G}})\\
\frac{\partial^2}{\partial\/r_i\,\partial\/r_j}\/n({\bf r})&=&-\sum_{\tilde{\bf G}} 
\tilde{G}_i\/\tilde{G}_j\,{\rm e}^{\immath\/{\tilde{\bf G}}\cdot {\bf r}}\,n({\tilde{\bf G}})\quad.
\end{eqnarray}
The potential due to a surface dipole in an orthorhombic cell with the surface in the $xy$-plane, 
i.e. lattice vector
${\bf a_3}= a_3\,{\bf e}_z$, is given by: 
\begin{equation}
\label{eq:vdipole}
  V^{\rm dipole}({\bf r})=(z\,\theta(z^{\rm min}-z)+(z-z^{\rm min})\,\theta(z-z^{\rm
    min}))\/{E}^{\rm field}+V_{0,{\rm dipole}}\quad,
\end{equation}
with the surface dipole moment $d$ according to Eq.~(\ref{eq:dipol}) and where
$z^{\rm min}$ is the coordinate at which $n^{\rm ave}(z)=\int d\/x\/\/d\/y\,\/n({\bf
  r})$ has its minimum. The surface dipole moment is given by
\begin{equation}
\label{eq:dipol}
d=d^{\rm el}-d^{\rm ion}\quad,
\end{equation}
where the electric dipole is
\[
d^{\rm el}= \int_{z^{\rm min}}^{z^{\rm min}+a_3}d\/z\int d\/x\/\/d\/y\,\/n({\bf r})
\]
and the ionic dipole is
\[
d^{\rm ion}=\sum_{I_{\rm s}, I_{\rm a}}\,q_{I_{\rm
    s}}\left(\boldsymbol{\tau}_{I_{\rm s}, I_{\rm a}}\cdot {\bf e}_z-z^{\rm
  min}+\theta(z^{\rm min}-\boldsymbol{\tau}_{I_{\rm s}, I_{\rm a}}\cdot {\bf
  e}_z)-a_3\right)
\]
with $V_{0,{\rm dipole}}$
\[
V_{0,{\rm dipole}}=-E^{\rm field}\left(\frac{d^{\rm ion}}{n^{\rm
    el}}-(a_{3}-z^{\rm min})\right)\quad,
\]
and where $E^{\rm field}$ is
\begin{equation}
\label{eq:field}
E^{\rm field}=-\frac{4\/\pi}{\Omega}\,d\quad.
\end{equation}
The Energy expressions evaluated in {\tt vofrho} read:
\begin{eqnarray}
\label{eq:etot}
E=E^{\rm kin}+E^{\rm H}+E^{\rm sr}-E^{\rm self}&+&\tilde{E}^{\rm ps,local}\nonumber\\&&
+E^{\rm ps,nl}+E^{\rm XC}+E^{\rm dipole} 
\end{eqnarray}
and
\begin{eqnarray}
E^{\rm Harris}=\sum_{\bf k}\/\sum_{i}&&\,w_{\bf k}\/f_{i, {\bf k}}\,
\epsilon_{i}({\bf k})-{\rm E}^{\rm H, el}[n^{(\tau-1)}]+E^{\rm
  XC}[n^{(\tau-1)}]\nonumber\\&-&\tilde{V}^{\rm XC}[n^{(\tau-1)}]
+E^{\rm H, Gau\ss}+E^{\rm sr}-E^{\rm self}-E^{\rm dipole}
\end{eqnarray}
with $n^{(\tau-1)}({\bf r})$ being the charge density of the previous
iteration and $T^{\rm s}$, $E^{\rm sr}$, $E^{\rm self}$ and $E^{\rm nl}$ 
after Eqn. (\ref{eq:ekin}), (\ref{eq:esr}), (\ref{eq:eself}) and (\ref{eq:enl}) rsp.\\
The Hartree energy is given by:
\begin{equation}
E^{\rm H}=2\/\pi\,\Omega\,\sum_{\tilde{\bf G}\neq 0}\,\frac{|n({\tilde{\bf G}})+n^{\rm Gau\ss}({\tilde{\bf G}})|^2}
{|{\tilde{\bf G}}|^2}
\end{equation}
The local pseudopotential energy is obtained from:
\begin{equation}
\tilde{E}^{\rm ps,local}=\Omega\,\sum_{\tilde{\bf G}}\,V^{\rm ps,local}({\tilde{\bf G}})\,
\overline{n({\tilde{\bf G}})}
\end{equation}
The exchange-correlation energy reads:
\begin{equation}
E^{\rm XC}= \int_{\Omega} d^3r\,\left(n({\bf r})+\tilde{n}^{\rm core}({\bf
r})\right)\,\epsilon^{\rm XC}\left(n({\bf r})+\tilde{n}^{\rm core}({\bf r})
\right)
\end{equation}
and the exchange-correlation potential energy is
\begin{equation}
\tilde{V}^{\rm XC}= \int_{\Omega} d^3r\,n({\bf r})\,V^{\rm XC}\left(n({\bf
  r})+\tilde{n}^{\rm core}({\bf r}) \right)\quad,
\end{equation}
where $\epsilon^{\rm XC}\left(n({\bf r})\right)$ is approximated by
LDA, BP, or PW91 (c.f. $V^{\rm XC}$ Eq.~(\ref{eq:vxc})).\\
Energy of the surface dipole $E^{\rm dipole}$ is:
\[
E^{\rm dipol}=-E^{\rm field}\,d
\]
with the dipole moment $d$ and the electrostatic
field $E^{\rm field}$ according to Eqn. (\ref{eq:dipol}) and (\ref{eq:field}) rsp.\\
The contributions to Harris energy $E^{\rm Harris}$ read:
\[
E^{\rm H, el}=2\/\pi\,\Omega\,\sum_{\tilde{\bf G}\neq 0}\,\frac{|n({\tilde{\bf G}})|^2}
{|{\tilde{\bf G}}|^2}
\]
\[
E^{\rm H, Gau\ss}=2\/\pi\,\Omega\,\sum_{\tilde{\bf G}\neq 0}\,\frac{|n^{\rm Gau\ss}({\tilde{\bf G}})|^2}
{|{\tilde{\bf G}}|^2}
\]
The local and ionic contribution to forces -- variable {\it fion} -- are given
by:
\begin{eqnarray}
{\bf F}_{I_{\rm s},{I_{\rm a}}}={\bf F}^{\rm sr}_{I_{\rm s},I_{\rm a}}-\,\immath\Omega\,
\sum_{\tilde{\bf G}}\,{\tilde{\bf G}}{\rm e}^{\immath\/{\tilde{\bf G}}\cdot{\bf r}}
&\,&\left\{\frac{4\/\pi}{|{\tilde{\bf G}}|}\,\overline{\left(n({\tilde{\bf G}})+n^{\rm Gau\ss}({\tilde{\bf G}})\right)}
\/\Phi^{\rm Gau\ss}_{I_{\rm s}}({\tilde{\bf G}}) \nonumber \right.\\&&
+\Phi^{\rm ps}_{I_{\rm s}}({\tilde{\bf G}})\,\overline{n({\tilde{\bf G}})}
\\&&\left.+\Phi^{\rm core}_{I_{\rm s}}({\tilde{\bf G}})\/\overline{V^{\rm XC}({\tilde{\bf G}})}{\/}^{\/}\right\}\quad,
\nonumber
\label{eq:fion}
\end{eqnarray}
with ${\bf F}^{\rm sr}_{I_{\rm s},I_{\rm a}}$ according to Eq.~(\ref{eq:fion_sr}).
\paragraph{Routine {\tt nlrhkb}}calculates the non-local contributions to
energy, potential and forces: 

The non-local contribution to energy $E^{\rm nl}$ -- variable {\it enl} -- is
given by:\\
\begin{equation}
E^{\rm nl}=\sum_{I_{\rm s},I_{\rm a}}\/\sum_{\bf k}\/\sum_{l,m}\/\sum_{i}\/w_{\bf k}
\/f_{i, {\bf k}}\/w^{\rm nl}_{I_{\rm s},l}\/|f^{\rm nl}_{i,I_{\rm s},I_{\rm a},l,m}(\bf k)|^2\quad,
\label{eq:enl}
\end{equation}
with $w^{\rm nl}_{I_{\rm s},l}$ and $f^{\rm nl}_{i,I_{\rm s},I_{\rm
    a},l,m}({\bf k})$ after Eq.~(\ref{eq:wnl}) and Eq.~(\ref{eq:fnl}) rsp.
If the nuclei are located on a mesh commensurate with the super cell the numerical
effort in evaluating (\ref{eq:enl}) can be reduced significantly as described
in~\cite{stumpf:94} -- see also Appendix~\ref{sec:app_tables}; the necessary
phase factors and other relevant quantities are tabulated by routine {\tt
  phfac}. 

The non-local contribution to forces ${\bf F}_{I_{\rm s},{I_{\rm a}}}$
--variable {\it fion\_nl}-- are computed from:\\
\begin{equation}
{\bf F}^{\rm nl}_{I_{\rm s},I_{\rm a}}= -\frac{2\/\pi}{a_{\rm lat}}\,
\sum_{\bf k}\/\sum_{l,m}\/\sum_{i}\/w_{\bf k}\/f_{i, {\bf k}}\,w^{\rm nl}_{I_{\rm s},l}\,
{\rm Im}\left\{\overline{{\bf f}_{i,I_{\rm s},I_{\rm a},l,m}({\bf k})}\/
f^{\rm nl}_{i,I_{\rm s},I_{\rm a},l,m}(\bf k)\right\}\quad,
\end{equation}
where $f^{\rm nl}_{i,I_{\rm s},I_{\rm a},l,m}({\bf k})$ and ${\bf f}_{i,I_{\rm s},I_{\rm a},l,m}({\bf k})$
--variables {\it fnl}, {\it sf1}, {\it sf2} and {\it sf3}-- are given by:\\
\begin{equation}
f^{\rm nl}_{i,I_{\rm s},I_{\rm a},l,m}({\bf k})=\sum_{\bf G} {\rm e}^{-\immath\/({\bf G}+{\bf k})
\cdot\tau_{I_{\rm s},I_{\rm a}}}\,\Phi^{\rm ps, nl}_{I_{\rm s},l,m}({\bf G}+{\bf k})\,
c_{i,{\bf G}+{\bf k}}
\end{equation}
and
\begin{equation}
\label{eq:fnl}
{\bf f}_{i,I_{\rm s},I_{\rm a},l,m}({\bf k})=\frac{a_{\rm lat}}{2\/\pi}
\sum_{\bf G} ({\bf G}+{\bf k})\/{\rm e}^{-\immath\/({\bf G}+{\bf k}){\bf \tau}_{I_{\rm s},I_{\rm a}}}
\Phi^{\rm ps, nl}_{I_{\rm s},l,m}({\bf G}+{\bf k})
\,c_{i,{\bf G}+{\bf k}}\quad,
\end{equation}
with $\Phi^{\rm ps, nl}_{I_{\rm s},l,m}({\bf G}+{\bf k})$ according to Eq.~(\ref{eq:pkg}).
\paragraph{Routine {\tt dforce}}computes the application of Hamiltonian to
wave function:
\begin{equation}
\langle {\bf G}+{\bf k}| \hat{H}^{\rm KS}|\Psi_{i,{\bf k}}\rangle
=\langle {\bf G}+{\bf k}| \hat{T}+ \hat{V}^{\rm local}
+ \hat{V}^{\rm ps, nl}|\Psi_{i,{\bf k}}\rangle\quad.
\end{equation}
The kinetic contribution reads:
\begin{equation}
\langle {\bf G}+{\bf k}| \hat{T}|\Psi_{i,{\bf k}}\rangle=\frac{1}{2}\,|{\bf G}+{\bf k}|^2
\,c_{i,{\bf G}+{\bf k}}\quad.
\end{equation}
The local potential contribution (obtained by FFT) is given by:
\begin{equation}
\langle {\bf G}+{\bf k}| \hat{V}^{\rm local}|\Psi_{i,{\bf k}}\rangle= \frac{1}{\Omega}\/
\int_{\Omega} d\/r^3\,V^{\rm local}({\bf r})\,u_{i,{\bf k}}({\bf r})
\,{\rm e}^{-\immath\/{\bf G}\cdot{\bf r}}\quad.
\end{equation}
The non-local potential contribution is obtained from:
\begin{equation}
\langle {\bf G}+{\bf k}| \hat{V}^{\rm ps, nl}|\Psi_{i,{\bf k}}\rangle=\sum_{I_{\rm s},I_{\rm a}}
\sum_{l,m}
\,w^{\rm nl}_{I_{\rm s},l}\/f^{\rm nl}_{i, I_{\rm s}, I_{\rm a}, l, m}({\bf k}) \,
{\rm e}^{\immath\/{\bf G}\cdot{{\bf \tau}_{I_{\rm s}, I_{\rm a}}}}
\Phi^{\rm ps, nl}_{I_{\rm s}, l, m}({\bf G}+{\bf k})
\end{equation}
with $w^{\rm nl}_{i_{\rm s},l}$ and $\Phi^{\rm ps, nl}_{I_{\rm s}, l, m}({\bf G}+{\bf k})$
according to (\ref{eq:wnl}) and (\ref{eq:pkg}) rsp. 

The same optimization as in routine {\tt nlrhkb} for nuclei located on a mesh commensurate with the
super cell applies here as well.
\subsection{Structure and form factors and ionic contributions to energy and
  forces}
\label{sec:app_form_struct}
\paragraph{Routine {\tt struct}}tabulates the structure factor of the ionic
basis --variable {\it struct}: 
\begin{equation}
S_{I_{\rm s}}({\bf G})=\sum_{I_{\rm a}}\,{\rm e}^{\immath\/{\bf G}\cdot{\bf \tau}_{I_{\rm s}, I_{\rm a}}}\quad.
\label{eq:struct}
\end{equation}
\paragraph{Routine {\tt formf}}tabulates the form factors of Gaussian charges
and of the pseudopotential and calculates the electrostatic self-energy of the
Gaussian charges:

The form factor of the local potential --variable {\it vps}-- is given by:
\begin{equation}
\Phi^{\rm ps}_{I_{\rm s}}({\bf G})=\frac{4\/\pi}{\Omega}\,\int_{0}^{\infty} d\/r\,r^2\,
j_0(r\/|{\bf G}|)\/\left\{V_{I_{\rm s},l_{\rm loc}}^{\rm ps}({\bf r})+\frac{q_{I_{\rm s}}}{r}\,
{\rm erf}\left(\frac{r}{r^{\rm Gau\ss}_{I_{\rm s}}}\right)\right\}\quad.
\label{eq:vps}
\end{equation}
The form factor of Gau\ss\ charges --variable {\it rhops}-- reads:
\begin{equation}
\Phi^{\rm Gau\ss}_{I_{\rm s}}({\bf G})=-\frac{q_{I_{\rm s}}}{\Omega}
\,{\rm e}^{-\frac{1}{4}\,{r^{\rm Gau\ss}_{I_{\rm s}}}^2\/|{\bf G|^2}}\quad.
\label{eq:rhops}
\end{equation}
The electrostatic self-energy of Gaussian charges --variable {\it eself}-- is
calculated from:
\begin{equation}
E^{\rm self}=\frac{1}{\sqrt{2\/\pi}}\/\sum_{I_{\rm s}}\,\frac{q_{I_{\rm s}}^2}
{r^{\rm Gau\ss}_{I_{\rm s}}}\/n^{\rm atom}_{I_{\rm s}}\quad.
\label{eq:eself}
\end{equation}

\paragraph{Routine {\tt nlskb}}tabulates the form factor and the prefactor of
non-local pseudopotentials:

The form factor of non-local pseudopotentials --variable {\rm pkg}, {\rm
pkg\_a}-- are obtained from:
\begin{equation}
\Phi^{\rm ps,nl}_{I_{\rm s},l,m}({\bf G})=\sqrt{\frac{4\/\pi}{2\/l+1}}\,\int^{\infty}_{0}
 d\/r \,r^2\,j_l(|{\bf G}|\/r)
\,\Delta V^{\rm ps,nl}_{I_{\rm s},l}(r)\,R_{I_{\rm s},l}(r)
\nonumber
y_{l\,m}(\theta_{\bf G},\phi_{\bf G})\quad,
\label{eq:pkg}
\end{equation}
where 
\[
\Delta V^{\rm ps,nl}_{I_{\rm s},l}(r)=V^{\rm ps}_{I_{\rm s},l}(r)-V^{\rm
  ps}_{I_{\rm s},l_{\rm loc}}(r)
\]
and $y_{l\,m}(\theta_{\bf G},\phi_{\bf G})$ are 
\begin{equation}
y_{l}^{m}(\theta,\phi) = \left\{ \begin{array}{ll}
                        \frac{1}{\sqrt{2}}\left( Y_{l}^{m}(\theta,\phi)+(-1)^{m}Y_{l}^{-m}
                        (\theta,\phi)\right) &\,\,\,\,\,\,\,,  m > 0\\
                        Y_{l}^{0}(\theta,\phi) &\,\,\,\,\,\,\,, m = 0 \\
                        \frac{1}{\sqrt{2}}\left( Y_{l}^{m}(\theta,\phi)-(-1)^{m}Y_{l}^{-m}
                         (\theta,\phi)\right)  &\,\,\,\,\,\,\,, m < 0
                       \end{array} \right.\quad.
\end{equation}

The prefactor $w^{\rm nl}_{I_{\rm s},l}$ --variable {\it wnl}-- reads:
\begin{equation}
w^{\rm nl}_{I_{\rm s},l}=\frac{4\/\pi}{\Omega}\/(2\/l+1)\left\{\int^{\infty}_{0}d\/r\,r^2\, 
R_{l,I_{\rm s}}(r)\,\Delta V^{\rm ps,nl}_{I_{\rm s},l}(r) \, R_{I_{\rm s},l}(r)\right\}^{-1}\quad.
\label{eq:wnl}
\end{equation}
\paragraph{Routine {\tt ewald}}calculates the screened ionic contributions to
the energy and the forces:

The energy of screened ions --variable {\it esr}-- is given by:
\begin{equation}
E^{\rm sr}=\sum_{\bf R}\/\sum_{I_{\rm s},I_{\rm a}}
\sum_{J_{\rm s},J_{\rm a}}\,
\frac{q_{I_{\rm s}}\/q_{J_{\rm s}}}{| \tau_{I_{\rm s},I_{\rm a}}-\tau_{J_{\rm s},J_{\rm a}}
-{\bf R}|}\,{\rm erfc}\left(\frac{| \tau_{I_{\rm s},I_{\rm a}}-\tau_{J_{\rm s},J_{\rm a}}-{\bf R}|}
{\sqrt{{r^{\rm Gau\ss}_{I_{\rm s}}}^2+{r^{\rm Gau\ss}_{J_{\rm s}}}^2}}\right)\quad,
\label{eq:esr}
\end{equation}
where in the innermost sum $(0,J_{\rm s},J_{\rm a})\neq ({\bf R},I_{\rm
  s},I_{\rm a})$ is obeyed. 
The forces due to screened nuclei-nuclei interaction --variable {\it
fion\_sr}-- is obtained from:
\begin{eqnarray}
{\bf F}^{\rm sr}=\sum_{\bf R}\/\sum_{I_{\rm s},I_{\rm a}}\sum_{J_{\rm s},J_{\rm a}}&&
q_{I_{\rm s}}\/q_{J_{\rm s}}\,\frac
{\tau_{I_{\rm s},I_{\rm a}}-\tau_{J_{\rm s},J_{\rm a}}-{\bf R}}
{| \tau_{I_{\rm s},I_{\rm a}}-\tau_{J_{\rm s},J_{\rm a}}-{\bf R}|^2} \nonumber \\
&\times&\left\{\frac{1}{\sqrt{\pi}}\,\frac{| \tau_{I_{\rm s},I_{\rm a}}-\tau_{J_{\rm s},J_{\rm a}}
-{\bf R}|}{\sqrt{{r^{\rm Gau\ss}_{I_{\rm s}}}^2+{r^{\rm Gau\ss}_{J_{\rm
        s}}}^2}}\,
{\rm exp}\left({-\frac{| \tau_{I_{\rm s},I_{\rm a}}-\tau_{J_{\rm s},J_{\rm a}}-{\bf R}|^2}
{{r^{\rm Gau\ss}_{I_{\rm s}}}^2+{r^{\rm Gau\ss}_{J_{\rm s}}}^2}}\right)\right.
\\&&\left. 
+\,{\rm erfc}\left(\frac{| \tau_{I_{\rm s},I_{\rm a}}-\tau_{J_{\rm s},J_{\rm a}}-{\bf R}|}
{\sqrt{{r^{\rm Gau\ss}_{I_{\rm s}}}^2+{r^{\rm Gau\ss}_{J_{\rm s}}}^2}}\right)\right\}\quad.
\nonumber
\label{eq:fion_sr}
\end{eqnarray}
where in the innermost sum $(0,J_{\rm s},J_{\rm a})\neq ({\bf R},I_{\rm
  s},I_{\rm a})$ is obeyed. The sum over ${\bf R}$ is  truncated at large $R^{\rm cut}$.

\section{The parameter file {\bf parameter.h} and the input file {\bf inp.ini}}
\label{sec:app_tables}
The parameter file {\file parameter.h} and the input file {\file inp.ini} are usually generated by the start utility {\file start} from the files {\file inp.mod} and 
{\file start.ini}. 
The program {\sf fhi96md} runs also individually without the help of the start utility
{\file start}. This requires the user to provide the files {\file parameter.h} 
and {\file inp.ini} in addition to the pseudopotential files and the optional restart files.

In Tables \ref{tab:parameter_h} and \ref{tab:inp_ini_1} we describe the parameter file 
{\file parameter.h} and the input file {\file inp.ini}. It should be noted that the file 
{\file  inp.mod} contains the entries {\it mesh\_accuracy} and {\it pfft\_store}, which have 
no effect in the program {\sf fhi96md} and are required only by the utility 
{\file start} to generate the parameters {\it nfft\_store} and {\it nr1x},
{\it nr2x}, and {\it nr3x}.

The file {\file parameter.h} is included by almost all source files at 
compilation time and is a requisite to compile the program {\sf fhi96md}. Consequently 
this file has to meet FORTRAN77 syntax rules and deviations from these may result in errors 
at compilation time. All of the parameters listed in Tab.~\ref{tab:parameter_h} have to be declared
as {\tt integer} variable and FORTRAN77 parameter statements are used to assign the corresponding 
values as described in this Table.
The parameter {\it ngwx} determines the maximum number of plane waves used to represent the 
wave function. For a given  energy cut-off $E_{\rm cut}$ (in Ry) it should be
set according to 
\[
ngwx \geq \frac{1}{6\/\pi^2}\,\Omega\,E_{\rm cut}^{\frac{3}{2}}\quad.
\]
{\it ngwix} for a given $E^{\rm init}_{\rm cut}$ should be set accordingly.
The size of the Fourier mesh determines the accuracy of the charge density. Set the
 parameters {\it nr1x}, {\it nr2x} and {\it nr3x} according to the sampling
 theorem: 
\[
  nr1x\geq \frac{2}{\pi}\,||{\bf a}_1||\,\sqrt{E_{\rm cut}}\quad,
\]
and {\it nr2x}, {\it nr3x} correspondingly. Using smaller values for {\it nr1x}, {\it nr2x} 
and {\it nr3x}   means to skip the highest ${\bf G}$-vectors in
\[
n({\bf r})=\sum_{\bf k}\sum_{i}\sum_{\tilde{\bf G}}\sum_{\bf G} w_{i,{\bf k}}
\overline{c_{i,\tilde{\bf G}+{\bf k}}}\,c_{i,{\bf G}+{\bf k}}\,
{\rm e}^{\immath\/({\bf G}-\tilde{\bf G})\cdot{\bf r}}
\]
and results in a better performance. However, the applicability of the grid
should be checked for each system individually. In particular in systems with
strongly localized orbitals this may be an unacceptable approximation.

The entries in file {\file inp.ini} are listed in Tab.~\ref{tab:inp_ini_1}. When separated by comma,
they are expected on the same line of file {\file inp.ini}. 
Whenever the entries have the same meaning as in the file {\file start.inp},
we refer the reader to Tab.~\ref{tab:start_inp_1}.

For each species {\it na} lines containing coordinates and the boolean parameters 
{\it tford} and {\it t\_auto\_coor} are expected. Lines containing the
velocity of a nucleus should immediately follow the line with the coordinates of the
corresponding ion.  Whenever some ions of a species are located on a mesh
commensurate with the super cell, the  evaluation of the non-local
contributions to energy and potential may be accomplished in a  more efficient
fashion. This requires the variable {\it tford} of the relevant ions to  
be set to {\it .false.} and the variable {\it t\_auto\_coord} to be set to {\it .true.}. The entry 
{\it ineq\_pos} contains the number of mesh points per super cell along each
lattice vector. Note, that the point (0,0,0) in the super cell must be a point
on the mesh.

\begin{ack}
The authors are indebted to Dr. E. Pehlke for many valuable discussions during
the development of this version of the package.
\end{ack}

\section*{TEST RUN}
\subsection*{Include file {\file inp.mod}}
\begin{verbatim}
-1  100 1000000              : nbeg   iprint timequeue
100 1                        : nomore nomore_init
12.0  0.2                    : delt  gamma
 4.0  0.2  0.0001            : delt2 gamma2 eps_chg_dlt
400 2                        : delt_ion nOrder
0.0 1.0                      : pfft_store mesh_accuracy
2 2                          : idyn i_edyn 
0 .false.                    : i_xc t_postc
.F. 0.001 .F. 0.002          : trane ampre tranp amprp
.false. .false. .false. 1800 : tfor tdyn tsdp nstepe 
.false.                      : tdipol
0.0001 0.0005 0.1            : epsel epsfor epsekinc
0.01 0.01 3                  : force_eps max_no_force
1                            : init_basis
\end{verbatim}
\subsection*{Include file {\file start.inp}}
\begin{verbatim}
 2                   : number of species (nsp)
 0                   : excess electrons 
 5                   : number of empty states
1  0                 : ibrav pgind     
10.47 0.0 0.0 0 0 0  : celldm 
 1                   : number of k-points
 0.5  0.5  0.5  1.0  : k-point coordintes, weight
 3 3 3               : fold parameter
 .false.             : t_kpoint_rel
 8  4.0              : Ecut [Ry], Ecuti [Ry]
 0.004 .true. .f.    : ekt tmetal tdegen
 .true. .false. 1    : tmold tband nrho
 5  2  1234          : npos  nthm  nseed
 873.0 1400.0 1e8 1  : T_ion T_init Q nfi_rescale
 .t. .true.          : tpsmesh coordwave
4  3 'Gallium'       : number of atoms, zv, name
1.0  3.0    0.7 3 3  : gauss radius, mass, damping,l_max,l_loc
.t. .t. .f.          : t_init_basis
0.0 0.0 0.0     .t.  : tau0 tford
0.5 0.5 0.0     .t.  : tau0 tford
0.5 0.0 0.5     .t.  : tau0 tford
0.0 0.5 0.5     .t.  : tau0 tford
4 5 'Arsenic'        : number of atoms, zv, name
1.0 3.0    0.7 3 3   : gauss radius, mass, damping, l_max,l_loc
.t. .t. .f.          : t_init_basis
0.25 0.25 0.25  .t.  : tau0 tford
0.75 0.25 0.75  .t.  : tau0 tford
0.75 0.75 0.25  .t.  : tau0 tford
0.25 0.75 0.75  .t.  : tau0 tford
\end{verbatim}
\subsection*{Include file {\file parameter.h} - generated by the start utility}
\begin{verbatim}
c========= Parameters for cp:================
      integer nsx,nax,nx,ngwx,ngx
      integer ngwix,nx_init,nr1x,nr2x,nr3x,nschltz
      integer nx_basis,max_basis_n,nlmax_init
      integer nnrx,nkptx,nlmax,mmaxx,n_fft_store
c SPECIES     ATOMS     STATES
      parameter(nsx= 2, nax=  4, nx= 21)
c FULL BASIS
      parameter(ngwx=   449, ngx=   3592)
c INITIAL BASIS
      parameter(ngwix=   168, nx_init=   169)
      parameter(nx_basis=     1, max_basis_n=    21)
      parameter(nlmax_init=     1)
c FFT:   X-MESH       Y-MESH       Z-MESH
      parameter(nr1x= 20,nr2x= 20,nr3x= 20)
      parameter(nnrx=     8400)
c K-POINTS
      parameter(nkptx=   4)
c Number of ffts to be stored between rhoofr and dforce
      parameter(n_fft_store=  1)
c Number of lm-components and max length of radial mesh
      parameter(nlmax= 4,mmaxx= 550)
      parameter( nschltz  =  1 )
c========= end of parameters for cp:==========
\end{verbatim}
\subsection*{Include file {\file inp.ini} - generated by the start utility}
\begin{verbatim}
   1   0         :ibrav, pgind
   32.0000    T    .00400    F         : nel,tmetal,ekt,tdegen
   8.00000   4.00000            : ecut,ecuti
    T    F  1       : tmold,tband,nrho
    5    2 1234         : npos, nthm, nseed
  873.00 1400.00   .1000E+09    1         : T_ion, T_init, Q, nfi_rescale
   2    T    T           :nsp,tpsmesh,coordwave
    4           : nkpt
    .1666667    .1666667    .1666667    .2962963 :xk(1-3),wkpt
    .1666667    .1666667    .5000000    .4444444 :xk(1-3),wkpt
    .1666667    .5000000    .5000000    .2222222 :xk(1-3),wkpt
    .5000000    .5000000    .5000000    .0370370 :xk(1-3),wkpt
   10.47000000     .00000000     .00000000      : lattice vector a1
     .00000000   10.47000000     .00000000      : lattice vector a2
     .00000000     .00000000   10.47000000      : lattice vector a3
    1.00000000     .00000000     .00000000      : rec. lattice vector b1
     .00000000    1.00000000     .00000000      : rec. lattice vector b2
     .00000000     .00000000    1.00000000      : rec. lattice vector b3
    10.4700000       1147.73082300         : alat,omega
'Gallium   '    4   3.00000  3.00000    : name,number,valence charge, ion_fac
  .70000 1.00000  3  3          : ion_damp,rgauss,l_max,l_loc
    T    T    F    :t_init_basis s,p,d
      .000000000      .000000000      .000000000    F    F    F    T
     5.235000000     5.235000000      .000000000    F    F    F    T
     5.235000000      .000000000     5.235000000    F    F    F    T
      .000000000     5.235000000     5.235000000    F    F    F    T
   0   0   0     : ineq_pos
'Arsenic   '    4   5.00000  3.00000    : name,number,valence charge, ion_fac
  .70000 1.00000  3  3          : ion_damp,rgauss,l_max,l_loc
    T    T    F    :t_init_basis s,p,d
     2.617500000     2.617500000     2.617500000    F    F    F    T
     7.852500000     2.617500000     7.852500000    F    F    F    T
     7.852500000     7.852500000     2.617500000    F    F    F    T
     2.617500000     7.852500000     7.852500000    F    F    F    T
   0   0   0     : ineq_pos
 24          : nrot = number of symmetries
  1---------
   1   0   0
   0   1   0
   0   0   1
  2---------
   1   0   0
   0   0   1
   0   1   0
........
 24---------
  -1   0   0
   0  -1   0
   0   0   1
\end{verbatim}
\subsection*{Output file file {\file fort.6}}
\begin{verbatim}
          ******* this is the complex fhi96md program ********
          *******          ibm - version              ********
          *******            Juli 1996                ********
>>>nbeg=  -1 nomore=    100  iprint= 100
>============================================
> Exchange: LDA 
>============================================
>>>electronic time step=  12.0000 gamma=    .2000
>>> Using delt=    4.0000 gamma=    .2000 when energy changes less than:   .1000E-03
>accuracy for convergency: epsel=  .00010 epsfor=  .00050 epsekinc=  .10000
 >damped Joannopoulos algorithm for electron dynamics
 >ions are not allowed to move
 normally no mixing of old charge is done
>ibrav=  1 pgind=  0 nrot= 24 alat= 10.470 omega= 1147.7308 mesh=   20   20   20
>ecut=    8.0 ryd   ecuti=    4.0 ryd    nkpt=   4
 nel, tmetal, ekt, tdegen= 32.0000000000000000 T 0.400000000000000008E-02 F
>alat=   10.470000 alat=   10.470000 omega= 1147.730823
 lattice vectors
a1   10.470000     .000000     .000000
a2     .000000   10.470000     .000000
a3     .000000     .000000   10.470000
 reciprocal lattice vectors
b1    1.000000     .000000     .000000
b2     .000000    1.000000     .000000
b3     .000000     .000000    1.000000
               positions tau0
 specie     Nr.     x         y         z
Gallium     1     .0000     .0000     .0000
Gallium     2    5.2350    5.2350     .0000
Gallium     3    5.2350     .0000    5.2350
Gallium     4     .0000    5.2350    5.2350
Arsenic     1    2.6175    2.6175    2.6175
Arsenic     2    7.8525    2.6175    7.8525
Arsenic     3    7.8525    7.8525    2.6175
Arsenic     4    2.6175    7.8525    7.8525
 nkpt= 4
 weight of all kpts: 0.999999900000000053
 ... so I'll scale them for you...
 ratios of FFT mesh dimensions to sampling theorem  1.061  1.061  1.061
 >ps-pots as given on radial mesh are used
>gvk: ngwx and max nr. of plane waves:     449     440
       k-point              weight    # of g-vectors
  1    .17    .17    .17   .0123      440
  2    .17    .17    .50   .0185      434
  3    .17    .50    .50   .0093      432
  4    .50    .50    .50   .0015      432
Weigthed number of plane waves npw:   435.248
Ratio of actual nr. of PWs to ideal nr.:    .99246
# of electrons=  32.0000, # of valencestates=  16, # of conduction states=  5
 atomic data for 2   atomic species
 pseudopotentialparameters for Gallium   
>nr. of atoms:  4, valence charge: 3.000,force fac:  3.00, speed damp: .7000
l_max  3 l_loc:  3 rad. of gaussian charge: 1.000
 pseudopotentialparameters for Arsenic   
>nr. of atoms:  4, valence charge:5.000,force fac:  3.00, speed damp:  .7000
l_max  3 l_loc:  3 rad. of gaussian charge: 1.000
 Final starting positions:
     .0000   .0000   .0000    5.2350  5.2350   .0000    5.2350   .0000  5.2350
     .0000  5.2350  5.2350    2.6175  2.6175  2.6175    7.8525  2.6175  7.8525
    7.8525  7.8525  2.6175    2.6175  7.8525  7.8525
\end{verbatim}
\begin{verbatim}
phfac: is, n_ideal: 1   0
phfac: is, n_ideal: 2   0
 phfac:is,i_kgv,n_class  1 0 0
 phfac:is,i_kgv,n_class  2 0 0
>nlskb:is=  1 wnl:  1   -.1067262  2   -.7541519  3   -.7541519  4   -.7541519
>nlskb:is=  2 wnl:  1   -.1008436  2   -.4304406  3   -.4304406  4   -.4304406
 --------------------------------------------
 starting density calculated from pseudo-atom
 --------------------------------------------
formfa: rho of atom in  3.000000
formfa: rho of atom in  5.000000
  
 =======================================================
                SYMMETRY OPERATIONS
 =======================================================
  
>s(isym) in latt. coord:          1        0        0          0        1 
       0       
   0        0        1
>sym(..) in cart. coord:    1.00000   .00000   .00000     .00000  1.00000  
 .00000  .00000   .00000  1.00000
>s(isym) in latt. coord:          1        0        0          0        0  
      1      0        1        0
........
 =======================================================
                 CHECK SYMMETRIES
 =======================================================
  
>Center of symmetry sym0=     .000000     .000000     .000000
 Table of symmetry relations of atoms,                     (iasym=nr. of
 symmetric at., xneu=tau0+<sym.Op.(isym)>)
   is  ia  iasym isym             tau0     xneu
   1   1     1     1      .00000   .00000   .00000      .00000   .00000  
 .00000
   1   1     1     2      .00000   .00000   .00000      .00000   .00000  
 .00000
........
   2   4     2    23     2.61750  7.85250  7.85250    -2.61750 -7.85250  
7.85250
   2   4     2    24     2.61750  7.85250  7.85250    -2.61750 -7.85250  
7.85250
  
 =======================================================
             ITERATIONS IN INIT STARTED
 =======================================================
........
 
 =======================================================
            ITERATIONS IN MAIN  STARTED
 =======================================================
  
 === LOOP n_it= 1
time elapsed for nlrh t =        .2900
 rhoofr stores starting density for mixing in c_fft_store

rhoofr: integrated electronic density in g-space = 31.999267 in r-space = 
31.999255
time elapsed for rhoofr t =       1.0000

  internal energy at zero temperature =   -33.989843 a.u.
             non-equillibrium entropy =      .000000 kB  
                 equillibrium entropy =      .000000 kB  
                            kT energy =      .004 eV  
                 (non-eq) free energy =   -33.989843 a.u.
                (non-eq) total energy =   -33.989843 a.u.
 \end{verbatim}
\begin{verbatim}
                        Harris energy =   -34.290291 a.u.
                       kinetic energy =    11.557750 a.u.
                 electrostatic energy =   -40.196264 a.u.
                  real hartree energy =     2.813098 a.u.
               pseudopotential energy =     6.191265 a.u.
           n-l pseudopotential energy =    -1.989500 a.u.
          exchange-correlation energy =    -9.553094 a.u.
exchange-correlation potential energy =   -12.456593 a.u.
          kohn - sham  orbital energy =    -2.810464 a.u.
                          self energy =    54.256150 a.u.
                           esr energy =      .000307 a.u.
                      gaussian energy =    22.685616 a.u.

 ===================================================
........
time elapsed for n x    nkpt x graham/ortho =        .1200
      nel    dampeig  true_efermi  efermi    ekt    seq     sneq
   32.0000      .700   3.20000   3.20000      .004    .00000    .00000

  k-point   .167  .167  .167, eigenvalues and occupation numbers:
eig  -9.053 -7.144 -7.144 -7.144 -3.100 -3.100 -3.100  -.402  -.402  -.402
eig    .557   .577   .577   .577  2.885  2.885  4.859  4.859  4.859  5.076
eig   6.576
occ  2.0000 2.0000 2.0000 2.0000 2.0000 2.0000 2.0000 2.0000 2.0000 2.0000
occ  2.0000 2.0000 2.0000 2.0000 2.0000 2.0000  .0000  .0000  .0000  .0000
occ   .0000
........
 >>>n_it nfi   Ekinc    Etot      Eharr    Ezero    mForce    mChange    
 Seq     Sneq   Efermi   Dvolt    W_a
>>>  1  0   4.67996  -33.98984  -34.29029  -33.98984    .00000     .000 
   .0000    .0000   3.2000    .0000    .0000
 >>> OK, I stop after timestep nr. 10000
time elapsed per electronic time step t =       3.3800
time in queue: 1000000 max. number of steps:  281060
 >>>fermi: No. of eigv with eig_force > 10% ekt: 84
>>>  2  0   7.31348  -34.26236  -33.98984  -34.26236    .00000     .000 
   .0000    .0000   3.2000    .0000    .0000
........
>>> 18  0    .06177  -34.35555  -34.35555  -34.35555    .00000     .000 
   .0000    .0000   3.2000    .0000    .0000
 === LOOP n_it= 19
phfac: is, n_ideal: 2   0

rhoofr: integrated electronic density in g-space = 32.000000 in r-space =
 31.999986

  internal energy at zero temperature =   -34.355551 a.u.
             non-equillibrium entropy =      .000000 kB  
                 equillibrium entropy =      .000000 kB  
                            kT energy =      .004 eV  
                 (non-eq) free energy =   -34.355551 a.u.
                (non-eq) total energy =   -34.355551 a.u.
                        Harris energy =   -34.355552 a.u.
                       kinetic energy =    11.920575 a.u.
                 electrostatic energy =   -40.142623 a.u.
                  real hartree energy =     2.909120 a.u.
               pseudopotential energy =     5.960148 a.u.
           n-l pseudopotential energy =    -2.496765 a.u.
          exchange-correlation energy =    -9.596886 a.u.
exchange-correlation potential energy =   -12.514237 a.u.
          kohn - sham  orbital energy =    -2.793552 a.u.
                          self energy =    54.256150 a.u.
                           esr energy =      .000307 a.u.
                      gaussian energy =    22.685616 a.u.

 ===================================================
 &&s atomic positions and local+nl forces on ions:
  Gallium   :
>&&s-n     .000000     .000000     .000000     .000000     .000000    
 .000000
>&&s-n    5.235000    5.235000     .000000     .000000     .000000     .000000
>&&s-n    5.235000     .000000    5.235000     .000000     .000000     .000000
>&&s-n     .000000    5.235000    5.235000     .000000     .000000     .000000
 \end{verbatim}
\begin{verbatim}
  Arsenic   :
>&&s-n    2.617500    2.617500    2.617500     .000000     .000000     .000000
>&&s-n    7.852500    2.617500    7.852500     .000000     .000000     .000000
>&&s-n    7.852500    7.852500    2.617500     .000000     .000000     .000000
>&&s-n    2.617500    7.852500    7.852500     .000000     .000000     .000000
>sum of all (local+nl) forces / n_atoms =     .0000000000    .0000000000 
   .0000000000 
(should = 0)
      nel    dampeig  true_efermi  efermi    ekt    seq     sneq
   32.0000      .700   3.20000   3.20000      .004    .00000    .00000

>  1. k-point   .167  .167  .167, ngw   440,        EWs and OCCs:
>eig: -9.370 -7.429 -7.429 -7.429 -3.427 -3.427 -3.427  -.549  -.549  
-.549   .260   .438   .438   .438  2.537
>eig:  2.537  4.705  4.705  4.705  4.847  6.349
>occ: 2.0000 2.0000 2.0000 2.0000 2.0000 2.0000 2.0000 2.0000 2.0000 
2.0000 2.0000 2.0000 2.0000 2.0000 2.0000
>occ: 2.0000  .0000  .0000  .0000  .0000  .0000
>  2. k-point   .167  .167  .500, ngw   434,        EWs and OCCs:
>eig: -8.621 -8.621 -7.350 -7.350 -3.535 -3.535 -1.754 -1.754  -.955  
-.955   .318   .318   .997   .997  1.277
>eig:  1.277  5.550  5.550  6.761  6.761  6.957
>occ: 2.0000 2.0000 2.0000 2.0000 2.0000 2.0000 2.0000 2.0000 2.0000 
2.0000 2.0000 2.0000 2.0000 2.0000 2.0000
>occ: 2.0000  .0000  .0000  .0000  .0000  .0000
>  3. k-point   .167  .500  .500, ngw   432,        EWs and OCCs:
>eig: -8.033 -8.032 -8.032 -8.032 -3.061 -3.061 -3.061 -3.061  -.039  
-.039  -.039  -.039  1.684  1.684  1.684
>eig:  1.684  5.608  5.608  5.608  5.608  7.767
>occ: 2.0000 2.0000 2.0000 2.0000 2.0000 2.0000 2.0000 2.0000 2.0000 
2.0000 2.0000 2.0000 2.0000 2.0000 2.0000
>occ: 2.0000  .0000  .0000  .0000  .0000  .0000
>  4. k-point   .500  .500  .500, ngw   432,        EWs and OCCs:
>eig: -7.951 -7.951 -7.951 -7.951 -3.815 -3.815 -3.815 -3.815  1.756  
1.756  1.756  1.756  1.756  1.756  1.756
>eig:  1.756  4.192  4.192  4.192  4.192  7.634
>occ: 2.0000 2.0000 2.0000 2.0000 2.0000 2.0000 2.0000 2.0000 2.0000 
2.0000 2.0000 2.0000 2.0000 2.0000 2.0000
>occ: 2.0000  .0000  .0000  .0000  .0000  .0000
 >>>n_it nfi   Ekinc    Etot      Eharr    Ezero    mForce    mChange 
    Seq     Sneq    Efermi   Dvolt    W_a
>>> 19  0    .05767  -34.35555  -34.35555  -34.35555    .00000     .000 
   .0000    .0000 3.2000    .0000    .0000
 ============= END OF THE MAIN-LOOP ================
av. time elapsed for nlrh t =        .2911
av. time elapsed for rhoofr t =        .9821
av. time elapsed for vofrho t =        .2632
av. time elapsed for n x nkpt x dforce =       1.4511
av. time elapsed for  nkpt x graham/ortho =        .1063
av. time elapsed for rest (in main) t =        .1979
av. time elapsed per elec. time step t =       3.2916
\end{verbatim}

\begin{figure}[h]
\epsfig{file=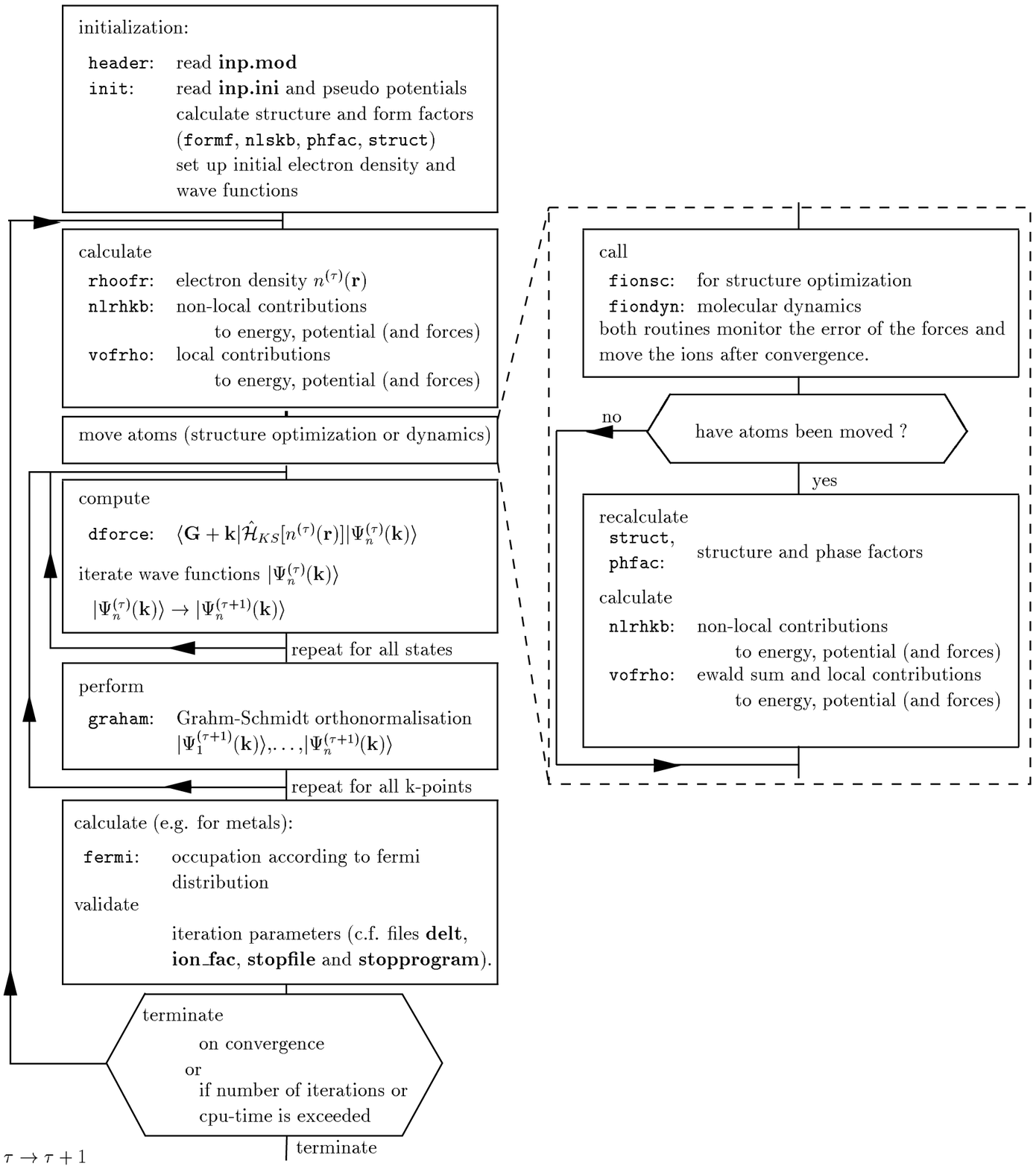,width=18cm}
\caption{\,}
\label{fig:flwchrt_main}
\end{figure}

\clearpage
\noindent Figure captions:
Fig~\ref{fig:flwchrt_main}:\parbox[t]{12cm}{ Flowchart of the program {\file fhi96md}. The routines {\tt init} and
  {\tt fiondyn} are described in more detail in the text. Output is generated
  at the end of each self-consistency cycle and by the routines {\tt fiondyn},
  {\tt fionsc}, {\tt fermi}, {\tt init} and {\tt vofrho}. Restart files are
  written by the routine {\tt fiondyn} and by a call of routine {\tt o\_wave}
  in the main program.}
\clearpage
\setcounter{section}{3}
\setcounter{table}{0}
\def\thesection      {\arabic{section}}
\renewcommand{\baselinestretch}{1.0}

\clearpage
{
\small
\tablecaption{input file {\file inp.mod}}
\label{tab:inp_mod_1}
\tablefirsthead{
\hline
\multicolumn{3}{c}{   inp.mod }\\
\hline 
parameter & type/range & explanation\\
\hline 
}
\tablehead{
\hline
\multicolumn{3}{c}{   inp.mod {\em continued}}\\
\hline 
parameter & type/range & explanation\\
\hline 
}
\tabletail{}
\begin{supertabular}{p{\la}p{\lb}p{\lc}}
\hline
{\it nbeg} &   & set up of the initial wavefunction: \\ 
& -1 &   $\psi_{\rm init}$ by an diagonalization on a subset of the plane wave
basis-set (c.f. Sec.~\ref{sec:dyn})\\ 
& -2 &     $\psi_{\rm init}$ is read from the file {\file fort.70}\\ \hline
{\it iprint} &integer  & number of electronic iterations between detailed output
and writing of restart files (c.f. text)\\ \hline
{\it timequeue}&  integer & maximum CPU-time in seconds:\\
               &              & the program terminates before exceeding this
                                time and restart files are written\\ \hline
{\it nomore} & integer & maximum number of :\\ 
       &  & if ({\it tfor}, {\it tdyn}=.false.): \mbox{self-consistency cycles}\\
&&if ({\it tfor} or {\it tdyn }=.true.): atomic moves\\ \hline
{\it nomore\_init}&integer&maximum number of self-consistency cycles in the initialisation\\\hline
{\it delt} & real & step length of the electronic iteration\\ \hline
{\it gamma}& real & if ({\it i\_edyn}=2): damping parameter \\\hline
{\it delt2}& real & second step length of electronic iteration (c.f. {\it eps\_chg\_dlt})\\\hline
{\it gamma2} & real & second damping parameter (c.f. {\it gamma} and {\it eps\_chg\_dlt})\\\hline
{\it eps\_chg\_dlt}& real & if total energy varies less than {\it
eps\_dlt\_chg}, {\it delt2} and {\it gamma2} replace {\it delt} and {\it
gamma} (is reset after moving nuclei)\\ \hline
{\it delt\_ion} & real &  molecular dynamics: time step for the integration of the equations of motion (a.u.) \\ \hline 
{\it nOrder}&  & if ({\it idyn}=0,1): order of the scheme:\\
&&  \tspc predictor corrector\\
& 0& \tspc \qquad 1 \qquad 2\\
& 1& \tspc \qquad 2 \qquad 3 \\
& 2& \tspc \qquad 3 \qquad 4\\
& 3& \tspc \qquad 4 \qquad 6\\ 
\hline
{\it pfft\_store}& real& percentage of wave functions for which a second transformation to real space is avoided (c.f. text)\\\hline 
{\it mesh\_accuracy}&real& degree to which the sampling theorem shall be satisfied (c.f. text)\\\hline
{\it idyn}&& scheme for solving the equation of motion of the nuclei:\\
& 0 &\tspc  predictor--corrector\\ 
& 1 &\tspc  predictor\\
& 2 &\tspc  Verlet\\
\hline
{\it i\_edyn} & & scheme to iterate the wave functions:\\
& 0 & \tspc steepest descent\\
& 1 & \tspc Williams--Soler algorithm\\ 
& 2 & \tspc damped Joannoupolos algorithm\\ \hline
{\it i\_xc} & & XC--functional:\\
& 0& \tspc LDA \mbox{(Cerperly/Alder, Perdew/Zunger)}\\
& 1 &\tspc Becke--Perdew XC (BP) \\
& 2 &\tspc Perdew (PW91) \\  \hline
{\it t\_postc} & logical  &\lindent{.true.:& post LDA with functional {\it i\_xc} \\
.false.:& start with functional {\it i\_xc}}\\ \hline
{\it trane} & logical&\lindent{.true.:&perturbe initial wave functions} \\ 
\hline {\it ampre} & real& if ({\it trane}=.true.): amplitude of random
perturbation added to wave function.\\ \hline {\it tranp} & logical &
\lindent{.true.:& perturbe positions of nuclei that are allowed to move
  (c.f. {\it tford} and {\it amprp})} \\ \hline {\it amprp}& real & if ({\it
  tranp}=.true.): amplitude of random perturbation of atomic positions (in
bohr)\\ \hline {\it tfor} & logical &\lindent{.true.:&structure optimization
  (c.f. {\it epsfor}, {\it force\_eps}, and, {\it tford}), set {\it
    tdyn}=.false. }\\ \hline {\it tdyn }& logical &\lindent{.true.:&molecular
  dynamics (c.f. {\it force\_eps} and {\it tford}), set {\it tfor}=.false.}\\ 
\hline {\it tsdp }& logical & structure optimization scheme:\\ 
&&\lindent{.true.:&modified steepest descent scheme\\ .false.:& damped
  dynamics scheme} \\ \hline {\it nstepe} & integer & if ({\it tfor} or {\it
  tdyn}=.true.): maximum number of electronic iterations allowed to converge
forces, the program terminates after {\it nstepe} iterations (c.f. {\it
  force\_eps})\\ \hline {\it tdipol}& logical & .true.: employ surface dipole
correction.\\ \hline & & the self-consistence cycle is terminated, if for the
last three iterations\\ {\it epsel} & real $>$ 0 & the variation of the total
energy is less than {\it epsel}\\ {\it epsekinc} & real $>$ 0 & and, if the
average change of wave functions is less than {\it epsekinc}\\ {\it epsfor}&
real $>$ 0 & and, if (tfor=.true.), forces on ions with {\it tford}=.true. are
smaller than {\it epsfor }\\ \\ \hline {\it force\_eps}& 2$\times$ real $>$ 0
& convergence criteria for local and total forces:\\ &&\cpindent{{\it
    force\_eps}(1):& maximum allowed relative variation in local forces
  before, if ({\it tfor}=.true.), executing a geometry optimization step or if
  ({\it tdyn}=.true.) calculating total forces.}\\ &&\cpindent{ {\it
    force\_eps}(2):& {maximum allowed relative} {variation in total forces}
  {before moving ions} ({\it tdyn}=.true.)}\\ \hline {\it max\_no\_force} &
integer & maximum number of electronic iterations for which no local forces
shall be calculated per atomic step\\ \hline {\it init\_basis}&&type of
basis-set of initialisation:\\ &1&plane-wave basis-set (c.f. {\it ecuti})\\ 
&2&LCAO basis-set (c.f. {\it tinit\_basis})\\ &3&mixed basis-set: LCAO and
plane waves\\ \hline
\end{supertabular}
}

\clearpage
\tablecaption{input file {\file start.inp}}
\label{tab:start_inp_1}
\tablefirsthead{
\hline
\multicolumn{3}{c}{  start.inp }\\
\hline
parameter & type/range & explanation\\
\hline 
}
\tablehead{
\hline
\multicolumn{3}{c}{  start.inp {\em continued}}\\
\hline
parameter & type/range & explanation\\
\hline 
}
\begin{supertabular}{p{\la}p{\lb}p{\lc}}
\hline
{\it nsp} & integer & number of atomic species\\ 
\hline
{\it nel\_exc}& real & number of excess electrons\\
\hline
{\it nempty } & integer $>$ 0 & number of empty states \\
\hline
{\it ibrav }& integer &
cell type (c.f. {\it celldm} and {\tt latgen.f}), e.g.:\\
&1& \tspc simple cubic lattice,\\ &2& \tspc fcc-lattice\\ &3& \tspc
bcc-lattice\\ &8& \tspc orthorhombic\\ \hline {\it pgind} & integer & point
group:\\ & 0 &\tspc automatic (symmetries and center)\\ & 1&\tspc no
symmetries\\ & \ldots&\tspc subgroup of lattice point group \\ &&\tspc (c.f.
{\it ibrav}, lib.sym and {\tt latgen.f} )\\ \hline {\it celldm(1..6)} &
6$\times$real & lattice parameters of the super cell (depends on {\it ibrav}).
Usually {\it celldm(1)} contains the lattice constant in bohr\\ \hline {\it
  nkpt} & integer & number of {k}-points\\ \hline {\it xk(1..3), wkpt}
&\parbox[t]{\lb}{{\it nkpts} lines of \\4$\times$real} & {\bf k}-points (c.f.
{\it t\_relative}) and weights of the {\bf k}-point, the sum over which must
be 1\\ \hline {\it i\_facs(1..3)}& 3$\times$integer $>$0&{\bf k}-point folding
factors (c.f. text) -- 1 1 1 : no folding\\ \hline {\it t\_kpoint\_rel} &
logical &\lindent{.true.:&frame of reference for {\bf k}-points is spanned by
  the reciprocal lattice vectors\\ .false.:&{\bf k}-points are given in
  Cartesian coordinates in units of $2\pi/a_{lat}$}\\ \hline {\it ecut} & real
& plane-wave energy cut-off (in Ryd)\\ \hline {\it ecuti} & real & plane-wave
energy cut-off of the initial wave function\\ \hline {\it ekt} & real &
temperature of the artificial Fermi smearing of the electrons in eV (c.f. {\it
  tmetal})\\ \hline {\it tmetal} & logical&occuppy electronic states according
to:\\ &&\lindent{.true.:&a Fermi distribution (c.f. {\it ekt})\\ .false.:&a
  step-like distribution}\\ \hline {\it tdegen} &
logical&\lindent{.true.:&occupation numbers are read from {\file inp.occ}
  (kept fixed for the run)}\\ \hline {\it tmold} & logical& \lindent{.false.:&
  only initialization is performed}\\ \hline {\it tband} &
logical&\lindent{.true.:&the electron density is not re-calculated after the
  initialization}\\\hline {\it nrho}& integer & set up of the initial electron
density:\\ & 1 &\tspc superposition of atomic electron densities\\ & 2 &\tspc
constructed from {\file fort.70}\\ & 3 &\tspc read in from {\file fort.72} \\ 
\hline {\it npos}&& if ({\it tdyn}=.true.): set up of initial positions of
nuclei and velocities ({\it tau0},{\it vau0}):\\ &1&\tspc {\it tau0, vau0}
from this file\\ &2&\tspc {\it tau0} from this file/{\file fort.70},{\it vau0}
from {\file fort.20}\\ &3&\tspc like No. 2, but {\it vau0}
according to {\it T\_init}\\ &4&\tspc like No. 1, but total momentum set to
zero\\ &5&\tspc like No. 3, but total momentum set to zero\\ &6&\tspc restart
from file {\file fort.20}\\ \hline {\it nthm}&&if ({\it tdyn}=.true.):
simulation ensemble (ions):\\ &0&\tspc microcanonical ensemble\\ &1&\tspc same
as 0, but rescaling of velocities\\ &2&\tspc canonical ensemble
(Nos\'e--Hoover)\\ &&(c.f. {\it nthm}, {\it nfi\_rescale} and {\it Q})\\ 
\hline {\it nseed}&integer&seed used to generate initial velocities \\ \hline
{\it T\_ion}&real&if (${\it nthm}=1,2$): temperature of nuclei in K \\ \hline {\it
  T\_init}&real&if (${\it npos}=3,5$): temperature of initial velocities in K
\\ \hline {\it Q}&real$>$0&if ({\it nthm}=2): mass of thermostat in a.u.  \\ 
\hline {\it nfi\_rescale}&integer$>$0&if ({\it nthm}=1): number of time steps
before velocities are rescaled \\ \hline {\it tpsmesh} & logical &pseudo
potentials are:\\ &&\lindent{.true.:&tabulated on logarithmic mesh\\ 
  .false.:&set up from parametrized form} \\ \hline {\it coordwave}&
logical&\lindent{.true.:&if ({\it nrho}=2), positions of nuclei are read from
  {\file fort.70}} \\ \hline \hline \multicolumn{3}{c}{{ {\it nsp} records
    describing each atomic species are expected below}}\\ \hline \hline
parameter & type/range & explanation\\ \hline {\it na}& integer & number of
ions of this species\\ \hline {\it zv}& integer & valence charge \\ \hline
{\it atom}& character*10 & name of the pseudo potential \\ \hline {\it
  rgauss}& real & radius of Gaussian charges\\ \hline {\it ion\_fac}& real$>$0
& if ({\it tfor}=.true.): mass parameter \\ &&if ({\it tdyn}=.true.): mass of
the nuclei in $[{\rm amu}]$ \\ \hline {\it ion\_damp} & 1$>$real$>$0 &if ({\it
  tfor}=.true. and {\it tsdp}= .false. ): damping parameter \\ \hline {\it
  l\_max}& 1,2,3 & highest angular momentum of the pseudopotential (1: s, 2:
p, 3: d) \\ \hline {\it l\_loc} & integer & angular momentum of the local
pseudo potential\\&$\le$ l\_max& \\ \hline {\it t\_init\_basis}&
3$\times$logical&\lindent{.true.:& include s,p and d orbital rsp. in LCAO
  basis-set (c.f. {\it init\_basis})}\\\hline {\it tau0(1..3)}, {\it tford}
&\parbox[t]{\lb}{\raggedright 3$\times$real, logical} & coordinates of the
nuclei (units depend on {\it ibrav}), flag whether nuclei may move\\ 

{\it vau0(1..3)} & 3$\times$ real & atomic velocities in a.u., expected only if ({\it tdyn}=.true. and  {\it npos}=1,4)
\\ \hline
\end{supertabular}
\clearpage
\renewcommand{\baselinestretch}{1.0}
\def\thesection{\Alph{section}}
\setcounter{section}{1}
\setcounter{table}{0}
\tablecaption{General symbols and the corresponding variable names.}
\label{tab:def}
\tablehead{}
\tablefirsthead{}
\tabletail{}
\begin{supertabular}{lll}
\hline
symbol&variable&\\
\hline
$n_{\rm sp}$&{\it nsp}& number of species\\
\hline
$n^{\rm atom}_{i_{\rm s}}$& {\it na(is)}& number of atoms per species\\
\hline
$n$&{\it n},{\it nx}& number of states\\
\hline
$n_{S}$&{\it nrot}& number of point symmetry operations\\
\hline
$l_{\rm loc}$&$l\_loc=l_{\rm loc}+1$& angular momentum of local pseudopotential\\
\hline
$l_{\rm max}$&$l\_max=l_{max}+1$&highest angular momentum\\
\hline
$s$&{\it s}(1-3,1-3,{\it irot})& point symmetry operation\\
\hline
$a_{\rm lat}$&{\it alat}&lattice constant\\\hline
$c_{i,{\bf k}+{\bf G}}$&{\it c0}({\it ig},{\it i},{\it ik})& wave function
coefficients\\
\hline
$f_{i,{\bf k}}$&{\it focc}({\it i},{\it ik})& occupation numbers\\
\hline
$w_{\bf k}$&{\it wik}& weights of {\bf k}-points\\
\hline
$\Omega$&{\it Omega}& volume of super cell\\\hline
$\tau_{i_{\rm s},i_{\rm a}}$&{\it tau0}(1-3,{\it ia},{\it is})
& ionic coordinates\\
\hline\\
\end{supertabular}

\clearpage
\tablecaption{Range over which the summation associated with the tabulated index is carried out.}
\label{tab:range}
\tablefirsthead{}
\tablehead{}
\tabletail{}
\begin{supertabular}{lcll}
\hline
index&range&variable&summation\\
\hline
$i_{\rm s}$&$1\le i_{\rm s}\le n_{\rm sp}$& {\it is}&species\\
\hline
$i_{\rm a}$&$1\le i_{\rm a}\le n^{\rm atom}_{i_{\rm s}}$& {\it ia}&
 ions of a species $i_{\rm s}$\\
\hline
$i$&$1\le i \le n$&{\it i}&electronic states\\
\hline
$l$& $0 \le l \le l_{\rm max}, l\neq l_{\rm loc}$&{\it i\_lm}& angular momentum\\
$m$& $ -l\le m \le l$&  &quantum numbers\\
\hline
${\bf G}$&$\frac{1}{2}||{\bf G}+{\bf k}||^2\le E_{\rm cut}$&{\it ig}&
reciprocal lattice vectors\\
${\tilde{\bf G}}$&$\frac{1}{2}||\tilde{\bf G}||^2\le 4\/E_{\rm cut}$&{\it ig}& \\
\hline
${\bf k}$& as given in input files &{\it ik}&{\bf k}-points\\
\hline
$s$& all point symmetries&{\it irot}&point symmetry
operations\\
\hline\\
\end{supertabular}

\clearpage
\setcounter{section}{2}
\setcounter{table}{0}
\tablecaption{parameter file {\file parameter.h}}
\label{tab:parameter_h}
\tablefirsthead{
\hline
\multicolumn{3}{c}{   parameter.h  }\\
\hline 
parameter & type/range & explanation\\
\hline 
}
\tablehead{
\hline
\multicolumn{3}{c}{   parameter.h  {\em continued}}\\
\hline 
parameter & type/range & explanation\\
\hline 
}
\tabletail{}
\begin{supertabular}{p{\la}p{\lb}p{\lc}}
\hline
{\it nsx}&integer& maximum number of ionic species\\ \hline
{\it nax}&integer& maximum number of ions per species\\ \hline
{\it nx}&integer& maximum number of electronic states per {\bf k}-point\\ \hline
{\it ngwx}&integer& maximum number of plane-waves, \\
{\it ngw}& 8$\times${\it ngwx}& maximum number of Fourier components of e.g. the electron density\\ \hline
{\it nx\_init}& integer& maximum number of basis functions in the initial
diagonalization\\ \hline
{\it ngwix}& integer& maximum number of plane waves in the initial
diagonalization\\ \hline
{\it nx\_basis}&integer&maximum number of atomic orbitals in the
initial diagonalization\\ \hline
{\it max\_basis\_n}&integer&max({\it nx},{\it nx\_basis})\\ \hline
{\it nlmax\_init}&integer& maximum number of atomic orbitals per atom in
initial diagonalization \\ \hline
{\it nr1x}, {\it nr2x} and {\it nr3x}& integer&size of the Fourier mesh (x, y and, z component rsp.)\\ 
{\it nnrx}& ({\it nr1x+1})$\times${\it nr2x}& number of mesh points \\
&$\times${\it nr3x}&\\ \hline
{\it nkptx}& integer& highest number of {\bf k}-points\\ \hline
{\it n\_fft\_store}&\parbox[t]{\lb}{\raggedright $1<{\rm integer}$}&number of
wave functions for which a second transformation to real space is avoided (c.f. text)\\ \hline
{\it nlmax}& integer &\parbox[t]{\lc}{$ nlmax\geq\sum_{l=0}^{l\_max-1} (2\,l+1) - 2l\_loc-1$}\\ \hline
{\it mmax}& integer& dimension of pseudo potential grid \\ \hline
{\it nschltz}& integer & \\
& 0 & \tspc scheme {\it i\_edyn}=2 disabled\\
& 1 & \tspc all schemes enabled\\ 
&& proper setting reduces storage needs\\ \hline
\end{supertabular}

\clearpage
\tablecaption{input file {\file inp.ini}}
\label{tab:inp_ini_1}
\tablefirsthead{
\hline
\multicolumn{3}{c}{   inp.ini  }\\
\hline 
parameter & type/range & explanation\\
\hline \hline
}
\tablehead{
\hline
\multicolumn{3}{c}{   inp.ini {\em continued}}\\
\hline 
parameter & type/range & explanation\\
\hline \hline
}
\tabletail{\hline}
\begin{supertabular}{p{\pdd}p{\pdd}p{\pbb}}
{\it ibrav}, {\it pgind}& 2$\times$integer& c.f. {\file start.inp}\\ \hline
{\it nel}, {\it tmetal}, {\it ekt} and, & real, logical, real&\lindentt{{\it nel}:& number of electrons}\\
{\it tdegen}&and, logical& c.f. {\file start.inp}\\ \hline
{\it ecut}, {\it ecuti}& 2$\times$real& c.f. {\file start.inp}\\ \hline
{\it tmold}, {\it tband}, {\it nrho}& 2$\times$logical, integer& c.f. {\file start.inp}\\ \hline
{\it npos}, {\it nthm}, {\it nseed}&3$\times$integer& c.f. {\file start.inp}\\ \hline
{\it T\_ion}, {\it T\_init}, {\it Q},&3$\times$real, integer&c.f. {\file start.inp}\\ {\it nfi\_rescale}&&\\ \hline
{\it nsp}, {\it tpsmesh},&integer,
logical&c.f. {\file start.inp}\\ {\it coordwave}&&\\\hline 
{\it nkpt}&integer&c.f. {\file start.inp}\\ \hline
{\it xk(1-3)}, {\it wkpt}&{\it nkpt} lines of& \\ &3$\times$real,
integer&k-points in Cartesian coordinates (units: $2\/\pi/a_{\rm lat}$)\\ \hline
{\it a1(1-3)}& 3$\times$real& lattice vectors in a.u. \\ 
{\it a2(1-3)}& 3$\times$real&  \\ 
{\it a3(1-3)}& 3$\times$real&  \\ \hline
{\it b1(1-3)}& 3$\times$real& reciprocal lattice vectors\\
{\it b2(1-3)}& 3$\times$real&  in $2\/\pi/a_{\rm lat}$ \\  
{\it b3(1-3)}& 3$\times$real&  \\ \hline
{\it alat}, {\it omega}&2$\times$real&lattice constant and cell volume in a.u.\\ \hline
\multicolumn{3}{c}{{\it nsp} records describing each ionic species (c.f. {\file start.inp})}\\ \hline
{\it name}, {\it na}, {\it zv}, {\it ion\_fac}&character*10,&\\ & 2$\times$integer, real&\\
{\it ion\_damp}, {\it rgauss},&2$\times$real, 2$\times$integer\\ {\it l\_max}, {\it l\_loc}&&\\
{\it t\_init\_basis}&&\\
\parbox[t]{\pdd}{tau0(1-3),\\t\_auto\_coord(1-3),\\tford}&\parbox[t]{\pdd}{3$\times$real,2$\times$logical}& c.f. {\file start.inp}\\&&{\it  t\_auto\_coord}: c.f. text\\
{\it vau0}(1-3)&3$\times$real& c.f. {\file start.inp}, expected only if ({\it npos}=1,4)\\
{\it ineq\_pos}&3$\times$integer&\\ \hline
{\it nrot}&integer& number of point symmetries\\ \hline
\multicolumn{3}{c}{{\it nrot} $3\times 3$ matrices representing point group elements }\\ \hline
s(irot)&\parbox[t]{\pdd}{3 lines of \\$3\times$integer}& point group element
preceded by a dummy line, the frame of reference is spanned by the lattice
vectors {\it a1}, {\it a2} and {\it a3}\\ \hline
\end{supertabular}
\end{document}